# Origin and Evolution of Saturn's Ring System

**Sébastien CHARNOZ**[*]
Université Paris Diderot/CEA/CNRS
Paris, France

**Luke DONES**
Southwest Research Institute
Colorado, USA

**Larry W. ESPOSITO**
University of Colorado
Colorado, USA

**Paul R. ESTRADA**
SETI Institute
California, USA

**Matthew M. HEDMAN**
Cornell University
New York, USA

(*): to whom correspondence should be addressed : charnoz@cea.fr

**ABSTRACT:**

The origin and long-term evolution of Saturn's rings is still an unsolved problem in modern planetary science. In this chapter we review the current state of our knowledge on this long-standing question for the main rings (A, Cassini Division, B, C), the F Ring, and the diffuse rings (E and G). During the Voyager era, models of evolutionary processes affecting the rings on long time scales (erosion, viscous spreading, accretion, ballistic transport, etc.) had suggested that Saturn's rings are not older than $10^8$ years. In addition, Saturn's large system of diffuse rings has been thought to be the result of material loss from one or more of Saturn's satellites. In the Cassini era, high spatial and spectral resolution data have allowed progress to be made on some of these questions. Discoveries such as the "propellers" in the A ring, the shape of ring-embedded moonlets, the clumps in the F Ring, and Enceladus' plume provide new constraints on evolutionary processes in Saturn's rings. At the same time, advances in numerical simulations over the last 20 years have opened the way to realistic models of the rings' fine scale structure, and progress in our understanding of the formation of the Solar System provides a better-defined historical context in which to understand ring formation. All these elements have important implications for the origin and long-term evolution of Saturn's rings. They strengthen the idea that Saturn's rings are very dynamical and rapidly evolving, while new arguments suggest that the rings could be older than previously believed, provided that they are regularly renewed. Key evolutionary processes, timescales and possible scenarios for the rings' origin are reviewed in the light of these recent advances.

# 1. Introduction

**1.1 New results on an old question**
Although Saturn's rings were first observed in the 17th century by Galileo Galilei, their origin and long-term evolution is still a matter of passionate debate. Whereas the origins of the rings of Jupiter, Uranus and Neptune, as well as of the dusty E and G Rings of Saturn, seem to be linked to the presence of nearby moonlets (via their destruction or surface erosion, see Esposito 1993; Colwell 1992; Burns et al., 2001; Hedman et al., 2007a; Porco et al., 2006), the unique characteristics of Saturn's main rings still challenge any scenario for their origin. Saturn's main rings have a mass on the order of one to several Mimas masses (Esposito et al., 1983, 2007; Stewart et al., 2007) and are mainly composed of pure water ice, with little contamination (Cuzzi and Estrada 1998; Poulet et al., 2003; Nicholson et al., 2008).

In the present chapter we detail the processes at work in the ring system, their associated timescales, and their possible implications for the origin and long term evolution of the rings. Meteoroid bombardment, viscous spreading and satellite perturbations imply a rapid evolution of the main ring system and support the idea of geologically young rings, although it seems very unlikely that the rings formed within the last billion years. Given the current state of knowledge, the destruction of a massive satellite (Pollack et al., 1976; Harris 1984) or the tidal splitting of comets grazing Saturn (Dones 1991; Dones et al., 2007) could be viable mechanisms for implanting such a large amount of ice inside Saturn's Roche zone. However, the actual cometary bombardment rate seems too low, by orders of magnitude, for these scenarios to work. Such a paradoxical situation could be resolved if the seemingly young rings are constantly renewed by some material recycling mechanism (sometimes called "cosmic recycling"). Developments in the last 20 years have shed new light on these ideas.

- *The Cassini spacecraft* has provided invaluable data that constrain particle properties, size distributions, and processes at work in the rings, such as: the discovery of propellers (Tiscareno et al., 2006, 2008), measurements of the shape of small inner satellites (Charnoz et al., 2007; Porco et al., 2007), multiple stellar occultations with resolutions better than 10 m (Colwell et al., 2006, 2007), high resolution spectra of rings (Nicholson et al., 2008), measurements of the rings' thermal inertia (see e.g., Leyrat et al., 2008) and the discovery of Enceladus' plumes (Porco et al., 2006).
- *Numerical simulations* have brought new insights into the microphysics and dynamics of ring particles below the kilometer scale, refining the notion of the Roche Limit and the conditions of accretion in a tidally dominated environment.
- *Solar System origin and evolution* is now much better understood than 20 years ago. In particular the discovery of the Kuiper Belt has brought important constraints on the primordial dynamical evolution of the giant planets, and our knowledge of satellite formation has also improved.

In this chapter we review how these three factors have modified and improved our conceptions regarding the origin and evolution of Saturn's rings.

**1.2 Organization of the chapter**
The challenge of writing such a chapter is that it requires a global, extensive study of ring physics and satellite dynamics, as well as planet and satellite formation. Since it is not possible to treat every aspect of these subjects in depth here, we will refer to other chapters of the present book in which several of these questions are addressed. Due to the large amount of material involved, we believe that a linear presentation is not desirable, as it would overwhelm a casual reader. Whereas the specialist will be interested in the details of some models, the graduate student will be more interested in the global picture. The present chapter was written to satisfy both readers.
Sections 2, 3 and 4 are mostly dedicated to Saturn's main rings (A, Cassini Division, B, and C). Section 2 is a self contained overview and should satisfy graduate students: it presents the main observations and the main evolutionary processes in a somewhat simplified way, as well as the associated timescales. This includes a discussion on the putative youth of the main ring system and a brief presentation of three possible scenarios for its origin.

Sections 3 and 4 are intended for specialists. Section 3 deals with physical models of evolutionary processes at work in the rings. It ends with a discussion on the long term evolution of Saturn's main rings. Section 4 describes in detail three scenarios for the main rings' origin, making use of recent results on satellite and planet formation. Since some material in section 2 is presented in more detail in sections 3 and 4, the discussion in section 2 will periodically refer to these sections.

Sections 5 and 6 describe the origins of Saturn's F Ring and the diffuse E and G Rings, respectively. While the origins and evolution of these rings are likely very different from those of the main rings, new *Cassini* data have revealed a variety of transient structures in the F Ring and have also provided new evidence connecting diffuse rings to material shed by satellites. A chapter on ring origins would not be complete without some discussion of these fainter rings.
In the conclusion we summarize what we have learned in the last 20 years. We also suggest future key observations that would help answer the long-standing question of the rings' origin and long term evolution.

## 2. Basic observational constraints and theoretical considerations

We review here the basic elements necessary to build a coherent model of the origin and evolution of Saturn's rings. The main physical characteristics of Saturn's rings that any formation and evolution scenario should account for are first recalled (section 2.1) and the main driving processes are reviewed in a simple way (section 2.2). Throughout this section, the reader will notice that there are still a lot of unknowns, both on the observational and theoretical side. Consequently we focus our attention on the rings' basic fundamental characteristics (material composition, mass, and spatial distribution). Only considering these three basic properties is already very challenging, as we will see. In section 2.3 we depict the basic scenarios for the origin of Saturn's massive ring system. Technical aspects of section 2 are described in detail in section 3 and 4, which are intended mainly for specialists.

**2.1 Ring structure**
Saturn's rings are the largest and the brightest of the four ring systems surrounding each of the giant planets (for comparison, see Esposito 2006). They likely contain at least as much mass as the moon Mimas (see section 2.1.3), and display all the phenomena found in the smaller ring systems of Jupiter, Uranus and Neptune. These phenomena include gaps with embedded moons and ringlets, narrow rings, broad rings, ethereal rings, waves, wakes and wiggles. Ring D lies interior to the brighter A, B and C Rings; Ring E is a broad, tenuous ring centered on the orbit of the moon Enceladus. The F Ring is a narrow ring just outside the A Ring, discovered by the Pioneer 11 flyby in 1979. The G Ring is another narrow ring outside ring F (Cassini entered the Saturn system in an apparently empty area between the F and G Rings in July 2004). Saturn's named rings are given in Table 1. Figure 1 compares the various rings, scaled to the equatorial radius of each planet. Saturn's rings include both massive and diffuse rings, and thus provide examples of phenomena occurring in all the ring systems. Because of the large mass of Saturn's rings, building a model of their origin may be the most difficult. However, any theory that explains Saturn's ring origin must be consistent with all the ring systems, and the main rings and faint rings are expected to have different origins and evolutions. Indeed, Saturn's rings are usually divided into two broad categories, main rings and diffuse rings, which display different physical properties and are driven by different physical processes:

- The *main ring system* containing rings A, Cassini Division, B, and C that are below Saturn's Roche limit for ice (about 140,000 km from Saturn's center, see section 2.2.1) and are made of particles larger than ~1 cm across. They are bright and collisionally evolved, with optical depths ranging from ~0.1 (C Ring, Cassini Division) to more than 5 (B Ring). Collisions are a major driving process for the main rings. Due to their closeness to Saturn tidal affects are also strong.

- The *faint ring system* includes the E and G Rings. They are mainly made of micrometer-sized dust. The E and G Rings are very faint and in general hardly visible from Earth. Due to their low densities, mutual collisions play almost no role in their evolution, although due to the small size of their constituent particles, non-gravitational forces (e.g., radiation pressure and Poynting-Robertson effects) are important to their evolution. By contrast, non-gravitational forces are generally assumed to play almost no role in the main ring system. Tidal effects play only a minor role in the E and G Rings because they are located outside the Roche limit for ice.

The F Ring is somewhat intermediate between the main rings and the faint rings because (1) it is located near the Roche limit and (2) it seems to include a mix of big bodies (larger than 1 meter, and perhaps as large as 1 km) surrounded by dusty transient structures comprised of micrometer-sized "dust". For more information about the structure of the main rings and faint rings, see the chapters by Colwell et al. and Horanyi et al., respectively.

**2.1.1 Ring particle sizes**
Saturn's rings are composed of myriad individual particles which continually collide. See the chapter by Schmidt et al. for a detailed discussion of ring dynamics. A growing consensus is that ring particles are actually agglomerates of smaller elements that are at least temporarily stuck together: these temporary bodies are subject to both growth and fragmentation. The balance between these competing processes yields a distribution of particle sizes and velocities (see sections 2.2.2 and 3.2). Like the ring systems of the other giant planets, Saturn's rings overlap with numerous small moons, including Pan and Daphnis. Not only do the nearby moons affect the rings dynamically, but they can also interchange material. These "ring moons" thus provide a source of material for making the rings, and also possible sinks, affecting the size distributions of particles.

The size of Saturn's ring particles extends over many decades, from fine dust to embedded moonlets, kilometers across (see chapter by Cuzzi et al.). The observations can often be fit with a size distribution following a power law

**$N(a)\, da = C_0\, a^{-q}\, da$ for $a_{min} < a < a_{max}$, Eq. 1**

where $N(a)da$ is the number of particles with radius between a and a+da, $C_0$ is a constant related to the total opacity, and $a_{min}$ and $a_{max}$ are the radii of the smallest and largest particles in the distribution. Typical values of q are around 3 for a<10m and range from 5 to 11 for a>10m (see Table 1 and Tiscareno et al., 2006; Sremčević et al., 2007 ). Note that q~3.5 is also characteristic of the asteroid belt and of size distributions created by shattering objects in the laboratory. Indeed, fragmentation processes at equilibrium tend to produce size distributions with q~3.5 (assuming size independent material strength, see Dohnanyi 1969). These similarities are likely not coincidental: both the asteroids and particles in planetary rings were probably created by fragmentation of larger objects and were subject to subsequent collisional evolution. However, collisions and limited accretion should both be at work in the rings, and the current size distribution is likely to be the result of these two competitive processes (see sections 2.2.4 and 3). Numerical simulations show that the collisions between particles tend to equalize the kinetic energy of the various particles, a process called the *equipartition* of energy. In ideal gases, this is achieved at the state of thermal equilibrium. For rings, this state is only partly reached: the smaller bodies have only 2% to 20% of the kinetic energy of motion of the largest (Salo 2001). However, because of their much smaller masses, the small ring particles have significantly higher velocities relative to a purely circular orbit. These larger velocities represent larger eccentricities and inclinations and are equivalent to a higher "temperature" that causes their vertical excursions also to be larger. Particles may sometime agglomerate in the form of larger extended structures (called "gravitational wakes", see chapter by Schmidt et al.) acting like big particles that may efficiently "heat up" small particles. Simulations also show that the size distribution should vary with distance: close to Saturn, tidal forces are strong so that large aggregates are prevented from forming (Albers and Spahn 2006) whereas close to the Roche Limit (around 140,000 km, see section 2.2) in the A Ring or F Ring for example, big rubble piles could exist.

The collisions of the ring particles can cause them either to grow in size or to be disrupted. The dynamic balance between these competing processes establishes an equilibrium state of aggregate bodies that resemble piles of rubble. Models show that particles tend to gather together, quickly growing to sizes that resist tidal disruption, only to be broken apart by mutual collisions (Weidenschilling et al., 1984a; Karjalainen and Salo 2004; Albers and Spahn 2006). Since relative velocities are low (about mm/s to cm/s) and collisions are inelastic, accretion is very rapid (see section 3.2). Large particles can hold smaller ones on their surfaces by their mutual gravitational attraction (Canup and Esposito 1995) or by adhesion (Albers and Spahn 2006). In Saturn's rings, the time scale is only weeks for house-sized objects to accrete. After rapid growth beyond several meters, ring particles become increasingly prone to disruption, and were therefore called "Dynamical Ephemeral Bodies" (Weidenschilling et al., 1984a). These large rubble piles are indeed dynamic and are very different from the simple idea of spherical ring particles of a uniform size. In the outer regions of Saturn's rings temporary aggregations are typically elongated and sheared, as seen in numerical simulations (e.g., Salo 1995; Lewis and Stewart 2005) and in various measurements (e.g., Colwell et al., 2007) of the self-gravity wakes in the A Ring, as well as one partially transparent F Ring feature ("Pywacket") observed by the Cassini UVIS and VIMS instruments (Esposito et al., 2008a).

**2.1.2 Ring particle composition**
Since spacecraft have not directly sampled the particles in Saturn's main rings, we must use reflection spectra and color to get some indication of their composition. In general, ring particles are similar to the nearby moons. Saturn's rings are predominantly water ice and therefore bright (Poulet et al., 2003, Nicholson et al., 2008); by comparison, Uranus's and Jupiter's are dark. Color variations across Saturn's rings may indicate varying composition, possibly due in part to the effects of the interplanetary dust that bombards them and darkens the particles. It is likely that Saturn's ring particles have rough, irregular surfaces resembling frost more than solid ice. There is good indication that the particles are under-dense (internal densities $\rho \ll 1$ g/cm$^3$), supporting the idea of ring particles as temporary rubble piles. These slowly spinning particles collide gently with collision velocities of just mm/sec. For more details on the spectroscopic signature of ring particles, see the chapter by Cuzzi et al.

**2.1.3 Mass of Saturn's rings**
The most striking characteristic of Saturn's rings is their vast extent and brightness: the rings of Jupiter, Uranus and Neptune are very faint, whereas Saturn's rings are among the most easily visible objects in the Solar System. Explaining this unique characteristic is also a major challenge for any formation scenario, and it is why measuring the mass and surface mass density of the rings is of major interest. Unfortunately we have very few data constraining the total mass of Saturn's rings. Pioneer 11 set an upper limit on the ring mass of $10^{-6}$ $M_{sat}$ ($M_{sat} \sim 5.7 \times 10^{26}$ kg stands for Saturn's mass) from the lack of any discernible perturbation on the spacecraft's trajectory (Null et al. 1981). Holberg et al. (1982) and Esposito et al. (1983) estimated the ring mass by using density waves to calculate the mass extinction coefficient $\kappa$ of the ring material from the measured density wave dispersion ($\kappa=\tau/\sigma$, with $\tau$ and $\sigma$ standing for the optical depth and the surface mass density, respectively). Esposito's measurements have now been confirmed by numerous measurements by Cassini in Saturn's A Ring (Tiscareno et al. 2007) and the Cassini Division (Colwell et al. 2008). Esposito et al. (1983) estimated the ring mass by assuming that this extinction coefficient measured in density wave regions applied everywhere in the rings, multiplying the optical depth measured by stellar occultation by the value $\kappa=1.3\pm0.5$ cm$^2$/g. Integration over the observed ring optical depth gives the ring mass $M_{ring}=5(\pm3)\times10^{-8}M_{sat}$. However, these results still do not measure the material in the densest parts of the rings, where most of the ring mass may reside. Esposito et al. (1983) noted that 40% of the calculated ring mass came from opaque regions: since no detectable starlight was observed in these regions, this gives only a lower limit. We now know that the measured opacity in the B Ring is mostly due to gaps between self-gravity wakes (e.g., Colwell et al. 2007). Neither the stellar or radio occultations from Cassini penetrate the densest regions. Colwell et al. (2007) set a lower limit on the wake optical depth of 4.9. No upper limit can be set. This means that even if the assumption of constant extinction coefficient is true for the B Ring, the integration cannot be trusted because the optical depth in much of the B Ring is still unmeasured. Dynamical calculations by Stewart et al. (2007) and Robbins et al.(2009) were used to predict the optical depth inferred from stellar occultations, given the clumpiness evident in dynamical simulations. For optically thick regions, the occultations would underestimate the amount of ring material by factors of 4 or greater. Using the model of Cooper et Al. (1985), estimating the ring density from secondaries created by galactic cosmic rays bombarding the rings, Esposito (2008a) shows that a non-uniform ring fits the data consistently with Cassini occultation results reported in Colwell et al. (2007). Thus, an alternate interpretation of the Pioneer 11 results is that the majority of the mass in the B Ring is in dense wakes. In that case, the measured secondary fluxes could be consistent with a total mass more than 5 times larger than originally reported by Cooper et al. for the B Ring.

## 2.2 Processes in Saturn's rings: simple considerations

We describe below the main physical processes that drive the evolution of Saturn's rings. We begin with a description of the notion of the Roche Limit, which seems to provide a natural answer to the simple and fundamental question: why are there rings rather than satellites close to Saturn (and the other giant planets)? Then we show that Saturn's rings are expected to evolve rapidly under the action of processes such as viscous spreading, surface darkening and material recycling. All these processes imply different evolution timescales, and raise the long–debated question: are the rings young or old? This is discussed in the last sub-section.

### 2.2.1 Tidal forces: the Roche Limit

Edouard Roche (1847; see Chandrasekhar 1969) calculated the orbital distance below which a purely fluid satellite would be pulled apart by tidal forces. This is the distance at which the gradient in a planet's gravitational force overcomes the gravitational attraction of the satellite's own material: its self-gravity alone is insufficient to hold it together. Of course, solid objects (and we humans, for example), can exist inside the Roche limit without being disrupted by a planet's tides due to their material strength. Even loose aggregates possess some strength. Roche's criterion can be written:

$$\frac{\alpha_r}{R} = 2.456\left(\frac{\rho_p}{\rho}\right)^{1/3}$$ **Eq. 2**

where fluid objects would suffer tidal disruption inside the Roche limit, $\alpha_r$, for a central planet with radius $R$ and average density $\rho_p$. The particle's density is $\rho$. Thus, more dense objects could avoid tidal disruption closer to the planet (smaller $\alpha_r$). For real bodies, stripping of loose material or fracture by tidal stresses occurs much closer to the planet than in the above equation. See Smoluchowski (1978, 1979) for discussion. The region surrounding the classical Roche limit, where the mutual gravitational attraction of two bodies is comparable to their mutual gravitation, is called the *Roche zone*. This is the same region where accretionary growth must compete with tidal disruption, so that formation of natural satellites around the planet would also be impeded there.

However, the definition of the Roche Limit is still ambiguous since its value depends sensitively on the underlying assumptions concerning the material strength. For example, Canup and Esposito (1995) numerically characterized accretion inside the classical Roche limit if one body is much larger than the other. In fact, if they are not rotating, a small body on the surface of a larger one will remain attached due to gravitational attraction if the distance to the planet is larger than a critical distance $\alpha$ such that:

$$\frac{\alpha}{R} \geq 1.26\left(\frac{\rho_p}{\rho}\right)^{1/3}$$ **Eq. 3**

So, considering Eqs. 2 and 3, the region between $\alpha$ and $\alpha_r$ could be considered as a (large) transition region in which some limited accretion could take place despite the tidal forces. Using Saturn's parameters and material density of water ice, this region extends from 70,000 to 136,000 km from Saturn's center, or about the totality of Saturn's main ring system. This opens the way to some possible material recycling through accretion processes (see sections 2.2.4 and 3).

### 2.2.2 Collisions: flattening and viscous spreading

A planetary ring consists of small particles in nearly circular orbits, with orbital angular velocities given approximately by Kepler's law,

$$\Omega = \sqrt{\frac{GM_{sat}}{r^3}}$$ **Eq. 4**

where $G$ is the gravitational constant, $r$ is the distance from the planet's center, and $\Omega$ is the angular rotation rate. The optical depth of a ring of equal-sized particles is

$$\tau = \pi a^2 \sigma / m$$ **Eq. 5**

(where $a$ and $m$ are the radius and the mass of an individual particle). For small values of $\tau$, the average number of collisions per orbit per particle is $\sim 2\tau$, since a particle will cross the ring plane twice. So the collision frequency is $\nu_c \sim \tau\Omega/\pi$. For an optically thick ($\tau > 1$) ring like Saturn's B Ring, collisions occur every few hours or less. This rapid collision rate explains why each ring is a nearly flat disk. Starting with a set of particle orbits on eccentric and mutually inclined orbits (say, the fragments of a small, shattered moon), collisions between particles dissipate energy:

they are not perfectly elastic, but also must conserve the overall angular momentum of the ensemble. Thus, the relative velocity is damped out, and the disk flattens after only a few collisions to a set of nearly coplanar, circular orbits (Brahic 1976).

Once the disk is flattened, it spreads radially, but on what timescale? Consider the mean free path $\lambda$ (average radial distance between collisions). For thick rings, this is the average random speed *c* multiplied by the time between collisions: $\lambda = c/(\tau\Omega)$. Cook and Franklin (1964) included both limiting values in their prescription (which was also adopted by Goldreich and Tremaine 1978a):

$$\lambda^2 = \frac{c^2}{\Omega^2}\frac{1}{1+\tau^2}$$  **Eq. 6**

The behavior of any individual particle experiencing repeated collisions can be seen as a simple random walk with the step size in radius given by $\lambda$. Let $\Delta r = n\lambda$. For a random walk, it takes on the average $n^2$ steps to reach a distance $n\lambda$ from the origin. Thus, the time for a typical particle to diffuse a distance $\Delta r$ is $n^2$ steps, each of duration $\Delta t = 1/(\Omega\tau)$, giving the total time for a particle to diffuse a distance $\Delta r$:

$$T \approx \frac{n^2}{\Omega\tau} = \left(\frac{\Delta r}{\lambda}\right)^2 \frac{1}{\Omega\tau} = \left(\frac{\Delta r}{c}\right)^2 \Omega\frac{1+\tau^2}{\tau}$$  **Eq. 7**

However, the particle velocity dispersion *c* is not an easily measurable quantity, while a related quantity known as the viscosity $\nu$ can be inferred from the observation of spiral density and bending waves (see the chapter by Schmidt et al.). It is therefore useful to consider an expression equivalent to Eq. 7, but using viscosity:

$$T \approx \frac{(\Delta r)^2}{\nu}$$  **Eq. 8**

Eqs. 7 and 8 show that the spreading timescale depends sensitively on $\nu$, *c* and $\tau$ (which depend on each other). Using Eq. 8 and viscosities reported in Tiscareno et al. (2007), we find that the A Ring doubles its width in 240 My to 2.4 Gy, a timescale shorter than the Solar System's age (using $\Delta r$ =15,000 km, $3\times10^{-3}$ m$^2$/s<$\nu$<$3\times10^{-2}$ m$^2$/s). Conversely for the B Ring, using Eq. 7 with $\Delta r$ = 25,000 km, $\tau$ = 3, c = $10^{-3}$ m/s (corresponding to a 10 m thick ring, as suggested by some numerical simulations, such as in Salo 1995), we find T ~ 12 Gy, longer than the age of the Solar System. For the C Ring, using Eq. 7, with $\Delta r$ = 20,000 km, $\nu$ = $2.7\times10^{-5}$ m$^2$/s (from the Daisaka et al., 2001 results for $\tau$=0.1) we find T ~ 700 Gy, much longer than the age of the Solar System. Thus there is a strong suspicion that the A Ring could be much younger than the B Ring (at least) by one or two orders of magnitude, and younger than the Solar System. Some theoretical results strengthen this argument. It has been shown that Eq. 7 does not hold for dense rings (Daisaka et al., 2001). The finite size and the gravity of the ring particles (e.g., enhanced transport by self-gravity wakes) are responsible for a global increase of the viscosity with the optical depth. Indeed, viscosity can be split into three different contributions (see chapter by Schmidt et al.): $\nu=\nu_{trans}+\nu_{coll}+\nu_{grav}$, corresponding respectively to the effect of particle transport ($\nu_{trans}$), the finite size of particles ($\nu_{coll}$), and the self-gravity of particles ($\nu_{grav}$). At low density, low optical depth and close to the planet, $\nu_{trans}$ and $\nu_{coll}$ dominate, consistent with Eq. 7. However, Daisaka et al. (2001) show that at higher densities and further away from the planet, random motions induced by the formation of wakes increase the values of $\nu_{trans}$ and $\nu_{grav}$, which become of the same order and increase rapidly with $\tau$. A fit to numerical simulations shows that $\nu(\tau)\propto\tau^\beta$ with $\beta\geq2$ (Daisaka et al., 2001). Eq. 8 shows that increasing the viscosity decreases the spreading timescale. As a consequence the A Ring, and perhaps the B Ring, could spread more rapidly than we have estimated here due to the presence of self-gravity wakes (Colwell et al., 2006, 2007).

**2.2.3 Meteoroid bombardment**
Like all Solar System objects, Saturn's rings suffer meteoroid bombardment. Due to their huge surface/volume ratio they are an efficient collecting area to capture the incoming meteoroid flux, which is strongly focused by the planet. The effect of meteoroid bombardment was studied in several papers (Ip 1984; Durisen et al., 1989, 1992, 1996; Cuzzi and Durisen 1990; Cuzzi and Estrada 1998) that highlighted its major role in the evolution of large-scale ring structures. When a meteorite hits the rings, typically at an impact speed larger than 30 km/s, the impact ejecta are thrown predominantly in the prograde orbital direction (Cuzzi and Durisen 1990) and spread radialy over hundreds to thousands of kilometers. These ejecta tend to be re-incorporated into the rings because they do not reach escape velocity from the ring system. The mass ratio of the ejecta to the impactor can be quite large (possibly up to $10^6$ depending on the impact geometry, material strength, etc.). All this results in a redistribution of material, as well as in a net transport of angular momentum across the system. Due to their higher surface/volume ratio, low-density rings such as the C Ring and the Cassini Division are more susceptible to being altered compositionally and dynamically. Under some assumptions on the meteoritic flux at Saturn, Cuzzi and Durisen (1990) found that the C Ring should spread into the planet in $10^7$-$10^8$ years due to the addition of mass from the meteoroids with very little net angular momentum. Note also that the C Ring could be regenerated by spillover of material from the B Ring on the same timescale, especially if the B Ring is more massive than thought (see section 3.1). The other major effect is a rapid darkening due to an accumulation of exogenic organic material. Over the age of the Solar System, Saturn's rings may have collected a mass comparable to their own mass, resulting in a strong darkening of the rings. Cuzzi and Estrada (1998) showed that the A and B ring's observed brightness implies that they are much younger than the age of the Solar System. They further suggest that the C to B ring transition is consistent with an exposure to meteoroid flux of around $5\times10^8$ years. Ballistic transport of material across the ring system may also create structures like sharp edges, wavy patterns, and optical depth ramps between regions of low to high optical depth, which recall many structures seen in the rings (Durisen et al., 1989, 1992, 1996). A more detailed discussion can be found in section 3.1.

Whereas meteoroid bombardment seems to be a major mechanism at play in Saturn rings, the exact magnitude of the bombarding flux is still poorly known. Attempts to observe the flash produced by 10 cm meteoroids onto the rings with the UVIS High-Speed Photometer (HSP) yielded no detections (but this is due to the low flux of light in the HSP bandpass for such impacts, Chambers et al., 2008). Measurements by Cassini's Cosmic Dust Analyzer are dominated by dust in the Saturn system, primarily the E Ring, and its cruise measurements did not allow for a determination of the micrometeoroid flux at Saturn (Srama et al., 2006; Altobelli 2006). The New Horizons dust experiment should provide a useful measure of the heliocentric variation in micrometeoroid flux on its way to Pluto (M. Horanyi, personal communication, 2006). Note however that the meteoritic flux at Jupiter was recently evaluated on the basis of Galileo data (Sremčevíc et al., 2005), and appeared to be in agreement, within a factor of 3, with previous estimates.

## 2.2.4 Cosmic Recycling

As shown in section 2.2.1, although the rings are located in Saturn's Roche zone, some limited accretion could be possible there, and accretion processes could substantially recycle material. Some evidence for this recycling can be found in Saturn's F Ring. Although the F Ring is clearly different from the main rings, the same processes of accretion and fragmentation occur there, and are more easily visible. If we can use the F Ring as an indicator of less obvious processes in Saturn's A and B Rings, this can provide a possible explanation of phenomena there. Cassini UVIS star occultations by the F Ring detect 13 events ranging from 27 m to 9 km in width (see Fig. 2). Esposito et al. 2008a interpret these structures as likely temporary aggregations of multiple smaller objects, which result from the balance between fragmentation and accretion processes. One of these features was simultaneously observed by VIMS and nicknamed "Pywacket". There is evidence that this feature is elongated in azimuth. Some features show sharp edges. At least one F Ring object is opaque, nicknamed "Mittens", and may be a "moonlet". F Ring structures and other youthful features detected by Cassini may result from ongoing destruction of small parent bodies in the rings and subsequent aggregation of the fragments. If so, the temporary aggregates are 10 times more abundant than the solid objects, according to Esposito et al. (2008a).

Calculations by Canup and Esposito (1995, 1997), Throop and Esposito (1998), and Barbara and Esposito (2002) (see Fig. 3) show that the balance between fragmentation and accretion leads to a bi-modal size distribution in the Roche zone, yielding a small number of larger bodies that coexist with the ring particles and dust. Thus, after disruption, some significant mass fraction of a shattered moonlet would be recaptured by other moonlets and would be available for producing future rings. Calculations by Barbara and Esposito (2002) show that at equilibrium after the disruption of a small moon, about 50% of the mass of small material is collected by the larger bodies, and the other half remains in the form of small fragments. This current loss rate implies no significant loss of mass from that ring over the age of the Solar System. These calculations show that recycling can extend the age of rings almost indefinitely, although this conclusion must be tempered by the results of Pöschel et al. (2007), which show that dissipation is an irreversible process and so processes such as viscous spreading cannot be reversed . If the recycling is large enough, this can significantly extend the ring lifetime (Esposito and Colwell, 2003, 2004, 2005; Esposito 2006). Larger bodies in the Roche zone can gradually grow if they can attain a roughly spherical shape (Ohtsuki et al., 2002). Those that do stick will continue to grow, perhaps until they are large enough to form a propeller structure (Spahn and Sremčević, 2000; Sremčević et al., 2002). The growth process, including many starts and stops, disruptions and rare events, may be very slow: Esposito et al. (2008a) call this "creeping" growth. Although adhesion and gravity alone may be insufficient to achieve km-size objects (Albers and Spahn 2006), solid cores (Charnoz et al. 2007) may persist to re-seed new agglomerates, or compaction may convert fragments into more competent objects. However, there was some suggestion in the past that moonlets could be present in the middle of the ring system (Spahn and Sponholz 1989), or next to its outer edge (see e.g. Spahn and Wiebicke 1989). Current numerical simulations do not include all the processes or long enough timescales to constrain this "creeping" growth.

## 2.2.5 Young or old rings?

The question of the rings' age is central to understanding their origin. Spreading of Saturn's A Ring due to mutual collisions among the particles (Esposito 1986; see section 2.2.2) and darkening of the rings due to infall and incorporation of meteoroid material (Doyle et al., 1989; Cuzzi 1995) both give ages shorter than the age of the Solar System (see section 3.1), from several $\times 10^5$ to several $\times 10^9$ years. Unless confined, the rings viscously spread as their particles exchange momentum via collisions and gravitational scattering (e.g. Goldreich and Tremaine 1982). Even if the rings are confined by shepherding satellites, the process only slows: the momentum is instead transferred to the shepherding moons via the resonance at the ring's edge. Because of the additional mass provided by the moon, the system's evolution is slower, but nonetheless the moon steadily moves away from the ring due to conservation of angular momentum (see section 3.4). The abundant density waves in the rings also transfer momentum to the perturbing moons, again causing the moons to recede. As an example, tiny Atlas, which orbits just outside Saturn's A Ring, would have evolved to its present location in less than 10 million years if it formed at the outer edge of the A Ring. Similar short timescales are found for Prometheus and Pandora, the F ring "shepherds", which could collide or be resonantly trapped in less than 20 million years (Poulet and Sicardy 2001). Mutual collisions and meteoroid bombardment grind the ring particles while charged particles sputter molecules from their surfaces.

Estimates of the age of the rings can discriminate between possible scenarios for ring formation. If the lifetimes of some ring features are much less than the age of the Solar System, those parts cannot have a primordial origin, but indicate the recent origin or possibly, renewal, of the material we observe. In Table 2 key observations and estimated ages are reported, and in particular the possibility that each feature could be renewed or not over time is reported. Inspecting Table 2 shows clearly that the rapid evolution of Saturn's rings argues for either a recent origin or renewal (or Cosmic Recycling, see section 2.2.4). However, if recycling seems an appealing possibility, Saturn's A Ring and the material in the Cassini Division must have been recently emplaced. Most of the ring mass is in the B Ring, where in much of the ring the optical depth is so large we cannot directly measure the mass from density waves. The B Ring could have survived longer and be less polluted now *if* its mass has been underestimated. As we will see, large uncertainties remain on the rings' age and processes at work.

At current rates of evolution, if the rings formed with Saturn, they would now be all gone or all dark (see Table 2). However, it is hard to reconcile the youthful aspects of the ring system with the large mass of Saturn's rings (perhaps greater than that of the moon Mimas). Although the rings of Jupiter, Uranus and Neptune are thought to be the result of the destruction of nearby kilometer-sized moonlets due to meteoritic bombardments (Esposito 1993; Colwell 1994) forming Saturn's rings would imply the destruction of a 200 km radius moon (!). An alternate explanation is the destruction of a close-passing comet (Dones 1991; Dones et al., 2007). As stated by several authors (e.g., Lissauer 1988, Ip 1988) such events are very rare and are unlikely to have occurred in the last billion years.

Ecliptic comets, which are generally thought to be the primary impactors on the giant planets and their regular satellites, are believed to originate in the so-called "Scattered Disk", a component of the Kuiper Belt. Since its discovery in 1993 (Jewitt and Luu 1993) the Kuiper Belt has been extensively observed and its dynamics has been simulated on computers. Colwell (1994) gives the number of cometary impacts ($N_i$) on a satellite (with radius $r_s$) with comets (radius $r_c$) :

$$N_i(r_s, r_c) = N_c(r_c) \cdot P \cdot \left(1 + \frac{1}{2}\left(\frac{V_{esc}}{V_\infty}\right)^2\right)\left(\frac{r_s + r_c}{1 km}\right)^2 \quad \text{Eq. 9}$$

where P is the intrinsic collision probability per year per km$^2$ with Saturn, $V_{esc}$ is the escape velocity from Saturn at the location of the satellite, and $V_\infty$ is the velocity at infinity of an ecliptic comet. $N_c(r_c)$ is the number of comets with size larger than $r_c$ in the cometary reservoir (here the Scattered Disk). Note that the ½ factor in front of the $(V_{esc}/V_\infty)^2$ is still a matter of debate (see Charnoz et al., 2009a). Levison et al. (2000) give $V_\infty \sim$ 3 km/s and find P $\sim 5.6 \times 10^{-24}$ km$^{-2}$/year in the present-day Solar System (also see Charnoz et al., 2009). Assuming a Mimas sized satellite ($r_s \sim$ 200 km) is present at 100,000 km from Saturn, a ~ 20 km radius comet is necessary to destroy such an object (Harris, 1984, Charnoz et al., 2009). Duncan and Levison (1997) estimate the current number of comets in the Scattered Disk with radii larger than 1 km is about 10$^9$. Assuming a size distribution at collisional equilibrium, so that $N_c(>r) \propto r^{-2.5}$, we get $N_c(>$ 20 km$) \sim 5 \times 10^5$. Using Eq. 9 with these values, we find that the number of destructive impacts on a ring progenitor located at 100,000 km from Saturn is about $N_i \sim 6.4 \times 10^{-12}$ per year (assuming the current dynamics and Kuiper Belt population). So the probability that such an impact happened in the last 10$^7$ or 10$^8$ years is only about $6 \times 10^{-5}$ or $6 \times 10^{-4}$. These numbers are so small that destruction of a ring parent body by a cometary impactor is unlikely to have happened in the last 10$^8$ years, as pointed out by several authors (Lissauer 1988; Ip 1988).

So we are left in a paradoxical situation where, on the one hand, evolutionary processes suggest that the rings are < 10$^8$ years old (or at least evolve on this timescale) and, on the other hand, cometary passages (either to break a satellite or to tidally disrupt a comet) are too rare to have happened recently in Solar System history.

For the moment, this paradox has no good solution. A possibility might be recycling (see discussion above) that may make the ring material look young despite having an age as old as the Solar System. Clearly, we do not really understand all the detailed processes at work in Saturn's rings and perhaps here lies the solution of this paradox.

## 2.3 An overview of possible scenarios for the origin of the main rings

Keeping in mind that we still do not understand how the rings could look so young if they formed shortly after Saturn's formation, we briefly summarize below several possibilities for ring formation, each of them with its pros and cons. The reader interested in details is invited to read section 4.

Because of the short time scales for viscous spreading of the accretion disk of the forming planet, gas drag, particle coagulation, and transport of momentum to the forming planet, Harris (1984) argues that rings did not form at the same time as their primary planets, but were created later by disruption of satellites whose large size had made them less subject to the early destructive processes. This could happen either at the end of accretion, or well after its completion. The pieces of the disrupted satellite are within the Roche zone, where tidal forces keep them from coagulating. This explains naturally the presence of shepherd satellites and ring moons around the giant planets as the largest pieces remaining after the destruction.

Another possibility is that Saturn's rings result from the same process that created the regular satellites. Like the ring particles, the satellites' orbits are prograde, nearly equatorial, and nearly circular. A question that immediately arises is whether rings are 1) the uncoagulated remnants of satellites that failed to form, or 2) the result of a disruption of a preexisting object, either a satellite or a comet. A related question highlighted by the apparent youth of the rings is whether this latter process of ring creation by satellite destruction continues to the present time. This possibility thus mixes the origin of the rings with their subsequent evolution. Whatever their origin, the sculpted nature of the rings of Saturn, Jupiter, Uranus, and Neptune requires active processes to maintain them.

*Are the main rings remnants from the Saturn's nebula?* Whereas the formation of giant planets is not fully understood, simulations show that during their formation giant planet cores (about 10-30 $M_\oplus$) are surrounded by a gaseous envelope which eventually collapses to form a compact disk. As the planet approaches its final mass, it forms an extended disk component (see Estrada et al. 2009; chapter by Johnson and Estrada). It is within this subnebula that the saturnian satellites form (Mosqueira and Estrada 2003a,b; Estrada et al., 2009). Some authors (Pollack et al, 1976) have suggested that today's rings are the unaccreted remnants of this disk. The main problem here is how this material may have survived long enough for the subnebula to dissipate. Gas drag could have easily swept all the ring-material into the planet's atmosphere. Another problem is the ring composition: why only ice would have survived, whereas we know there were silicates in this disk (from the density of Saturn's satellites). Whereas this scenario seems today the most unlikely, it has never been really investigated in detail. See discussion in section 4.1 for further details.

*Are the main rings remnants from a destroyed satellite?* Since the total ring mass is comparable to a Saturnian mid-sized satellite (like Mimas) it is tempting to suggest that a 200-300 km radius satellite was destroyed in Saturn's Roche Zone. The first question is: how to bring such a big object into Saturn's Roche Zone? One possibility (see section 4.2.1) is that it was brought there during Saturn's formation through type 1 migration in Saturn's subnebula or gas drag (Mosqueira and Estrada 2003a,b; chapter by Johnson and Estrada), in a similar fashion to "Hot Jupiters" in extrasolar systems. Once in the Roche Zone, the satellite must be destroyed by an impactor, since tidal forces alone cannot grind a satellite to 1 cm sized particles (see section 4.2 for more details). An appealing aspect of this scenario is that it has been recently shown (Charnoz et al., 2009) that a Mimas-mass moon located $10^5$ km from Saturn can be destroyed during a "Late Heavy Bombardment" (LHB) type event (Tsiganis et al., 2005) about 700 My after the planet's birth. Conversely, such a mechanism cannot work for Uranus and Neptune because tidal migration should have removed the satellite from their host planet's Roche Zone before the onset of the bombardment (Charnoz et al., 2009). This encouraging property may be a partial explanation for Saturn's having massive rings. However, the LHB hypothesis sets a ring age around 3.8 or 3.9 Gy, still much too old to be reconciled with the apparent youth of the rings. In addition, a satellite should be made of ice and silicates, and silicates seem nearly absent in Saturn's rings, in apparent contradiction with this scenario. See section 4.2 for further discussion of this issue.

*Are the main rings remnants from a tidally split comet?* Dones (1991) and Dones et al. (2007) suggested that a big (300 km radius or more) "comet" or Centaur could be the progenitor of Saturn's rings, provided it passed very close to Saturn and was disrupted by the tidal forces close to the planet, in a fashion very similar to how comet Shoemaker-Levy 9 was disrupted after a close passage with Jupiter. Again, the main problem is that the current flux of objects larger than 300 km is far too low, by orders of magnitude, for such a close passage to have been likely in the last $10^8$ years. However, this scenario was revisited in Charnoz et al. (2009), who considered the possibility that this happened during the LHB. Charnoz et al. (2009) show that the cometary flux is so high during this event that several tens of Mimas masses of cometary material may have been brought into Saturn's Hill sphere, and that a fraction of it could have ended in Saturn's Roche zone. Surprisingly, the same thing may have happened to other giant planets during the LHB. So explaining why only Saturn has massive rings is a mystery in this scenario. See section 4.3 for more details.

## 2.4 Beyond the paradox?

We have presented the main observations and the main ideas concerning processes affecting the evolution of Saturn's main rings and possible scenarios for their origin. We have reached a point of giving seemingly paradoxical conclusions (rings look young, whereas they cannot have formed recently). Such a situation arises from a lack of knowledge of the processes and of the underlying assumptions, such as material recycling and total ring mass. For example, recycling of ring material can not explain the limited micrometeoroid darkening of Saturn's rings (Cuzzi and Estrada 1998). Why are the rings not darker now, if they are truly ancient? One possibility is that the total mass of the rings, mostly in Saturn's B Ring, has been underestimated. As said in section 2.1.3, for example, Cooper et al. (1983) selected the smaller of two possible values for the B Ring mass consistent with Pioneer 11 results. Esposito (2008a) showed that the "granola bar" model of Colwell et al. (2007) is also consistent with Pioneer 11 results, and implies a B Ring mass about 4 times greater than estimated by Esposito et al. (1983). Because the total optical depth of the B Ring is still unmeasured and may be more than two times greater than previous estimates (Colwell et al., 2007; Stewart et al., 2007), meteoritic pollution would have a smaller effect. This is seen in the Markov chain simulations of Esposito et al. (2008b). An important consideration is that meteoritic pollution could affect mostly the exterior of objects (Esposito and Eliott 2007). If the rings (particularly the B Ring) are much more massive than we now estimate, the interior of the largest ring objects (which may encompass most of the ring mass) can remain more pristine until disrupted. However, in the thinner parts of the rings where density waves are visible, we have good ring mass estimates. The mass there would be quickly polluted by the micrometeoroid flux (Cuzzi and Estrada 1998). Esposito (1986) noted that most of the age problems involve Saturn's A Ring. Perhaps the A and F Rings are more recent? This raises the problem of how the material that formed these (possibly more recent) rings had been preserved, perhaps as large, unconsolidated objects with competent solid cores encased in rubble. If the A Ring is much younger than the B ring, we also need to find an explanation for the Cassini Division. It would not have originated simply by a density wave clearing a gap, as proposed by Goldreich and Tremaine (1978b).

From this discussion, it is clear that processes affecting the rings' evolution and scenarios for their origin must be investigated with much care in order to determine the limits of our ignorance. This is the subject of the following two sections.

# 3. Evolution of the main rings

This section details processes affecting the large scale evolution of the rings. It can be skipped by undergraduate students and is mainly intended for specialists. It illustrates the diversity and complexity of processes at work, whose interactions are still not really investigated. It also shows that the timescales of these processes, discussed in the previous sections, sometimes rely on some poorly constrained parameters (like the meteoritic flux, the B Ring surface mass density, etc.). Better understanding these complex processes will undoubtedly help to resolve the paradox discussed in the previous section.

Viscous spreading was treated in the previous section. Section 3.1 presents the effect of meteoritic bombardment. Section 3.2 presents the effects of accretion within the rings, which is a key element for the Cosmic Recycling process (and might be a solution for the apparent youth of the rings). Section 3.3 presents the effects of collisional cascades that counteract accretion processes. Section 3.4 presents the effects of ring-moon interactions that may give further clues on the rings' age and evolution. Section 3.5 is a tentative exploration of the future evolution of Saturn's ring system.

## 3.1 Meteoritic bombardment and ballistic transport

Because the rings have a huge surface area-to-mass ratio, they are particularly susceptible to modification due to extrinsic meteoroid bombardment. While other mechanisms can lead to the erosion of ring particle surfaces, such as interparticle collisions, sputtering of energetic ions or photons, and the sublimation of ices, most tend to produce atomic and molecular by-products, whereas meteoroid impacts can produce a large amount of particulate ejecta in addition to significant amounts of gas and plasma (e.g., see Morfill et al., 1983). Moreover, the vast majority of the dust and debris produced from these collisions are ejected at speeds much less than the velocity needed to escape the Saturn system at the distance of the rings (~27 km/s). As a result, a copious exchange of ejecta between different ring regions can occur, which over time can lead to the structural and compositional evolution of the rings on a global scale. This process by which the rings evolve subsequent to meteoroid bombardment is referred to as "ballistic transport" of impact ejecta (Ip 1983; Lissauer 1984; Durisen 1984a,b).

### 3.1.1 Principles

The essence of the ballistic transport mechanism is shown schematically in Fig. 4. Impact ejecta from a given (non-disruptive) meteorite impact are thrown predominantly in the prograde orbital direction. This results naturally from consideration of impact geometries and probabilities azimuthally averaged over the rings (Cuzzi and Durisen 1990). The yield $Y$ of a single impact, which is defined as the ratio of ejecta mass to impactor mass, can be quite large, depending on several factors. Oort cloud-type projectiles, for example, impact at speeds much greater than the sound speed in either the target ring particle or impactor. As a result, shocks vaporize part or the entire projectile along with a small volume of the target, and shatter and excavate a large volume of the target (e.g., Durisen 1984b). The velocity (or velocity distribution) at which it is ejected depends on the hardness of the target and the angle of impact (Cuzzi and Durisen 1990). If the target is powdery, yields can be on the order of ~ $10^5 - 10^6$ at normal incidence (Stöffler et al., 1975; Burns et al., 1984), while micrometer-sized particles impacting into granular surfaces have yields from 1 to $\geq 10^3$ (Vedder 1972, Koschny and Grün 2001). Ejecta velocities for the bulk of the material tend to range from ~ 1 meter/second to tens of meters/second, while escape velocities are achieved for a mass comparable to the projectile (see Durisen 1984b and references therein). This means that, in general, one need not consider the total mass added to or lost from the system unless one considers very long evolutionary times.

Impact ejecta carry not only mass, but angular momentum as well. Because most ejecta are prograde, it tends to land in the rings at larger distances where the specific angular momentum is greater, so the net resultant drift is inward. Ring structure (i.e., optical depth and surface mass density) can have an effect on the rate of material drift because the ejecta absorption probability (which determines the actual mass that hits the rings as opposed to merely passing through them) depends weakly on the local optical depth, but its angular momentum depends linearly on the surface density (Cuzzi and Durisen 1990; Cuzzi and Estrada 1998). Moreover, ring particles tend to be smaller in low optical depth regions. Thus, lower surface density regions are more quickly altered compositionally (e.g., darkened) relative to higher surface density regions simply because the mass fraction of extrinsic ("polluting") material relative to the local overall mass surface density (which is related to the optical depth) will be higher in low surface density regions, for a given micrometeoroid flux.

The value of the micrometeoroid flux at Saturn (now, and in the past where it was most likely greater) remains uncertain. Past estimates of the micrometeoroid flux at Saturn (Morfill et al., 1983; Ip 1984; Cuzzi and Durisen 1990; Cuzzi and Estrada 1998) vary slightly but all imply that the main rings would be impacted by close to their own mass over the age of the Solar System (Landgraf et al., 2000). More recently, Galileo measurements of the flux at Jupiter have provided estimates of the mass flux that may at most be too low by a factor of 2 – 3 only (Sremčević et al., 2005) compared to previous estimates. However, the micrometeorite mass flux at Saturn has not been observed, and will not be until the Cassini Extended Mission, which is now underway, using an indirect technique similar to that described in Sremčević et al. (2005).

### 3.1.2 Dynamical evolution

For nominal values of ring mass and meteoroid flux, micrometeorite impacts on the rings will have two different effects. First, they will lead to angular momentum loss which would lead, for example, to the loss of the C Ring to the planet in $\sim 10^7 - 10^8$ years (Cuzzi and Durisen 1990). A similar age for the rings more closely related to ejected material was obtained by Northrop and Connerney (1987), who suggested that water molecules generated by impacts were lost to Saturn's ionosphere, providing an explanation for Saturn's unusually low ionospheric electron density. Note that resonant interactions with nearby "ring moons" can also lead to significant angular momentum loss by the rings (Goldreich and Tremaine 1982; Poulet and Sicardy 2001). Second, in addition to angular momentum arguments, meteoroid material also darkens and pollutes the rings over time. Doyle et al. (1989) and Cuzzi and Estrada (1998) noted that the relatively high albedo of the A and B rings was inconsistent with their having retained more a small fraction of primitive, carbonaceous material from the large mass they would have accreted over the age of the Solar System (using a mass flux from Morfill et al., 1983), thereby suggesting a geologically young age for the rings.

The first efforts to model the global effects of ballistic transport on Saturn's rings indicated that ballistic transport can produce prominent edge effects, enhance pre-existing density contrasts, and produce "spillovers" from high to low optical depth regions (Ip 1983; Durisen 1984a). Since then, the influence of meteoroid bombardment and ballistic transport has been found to explain certain aspects of ring structure such as the fairly abrupt, and remarkably similar, inner edges of the A and B Rings, which include "ramp" features which connect them to the Cassini division and C Ring, respectively (Durisen et al., 1989, 1992, 1996), given evolutionary times of ~ 100 "gross erosion" times (~ $10^8$ years, see Fig. 5). A gross erosion time is defined as the time a reference ring annulus would disappear due to ejected material if nothing returned. Durisen et al. (1992) concluded from their simulations that the sharpness of the A and B Ring inner edges are maintained for very long times at their observed widths by a balance between ballistic transport, which tends to sharpen low-optical-depth to high-optical-depth transitions, versus the broadening effect that results from viscous transport. In addition, Durisen and colleagues found that prograde ejecta distributions (Cuzzi and Durisen 1990) inevitably lead to the formation of a ramp on the low optical depth side of the ring edge due to advective effects.

Another effect unique to ballistic transport found by Durisen et al. (1992) is the formation of wavelike structure that tend to be propagate away from edges (see Fig. 5), a phenomenon that Durisen et al. called "echoing". The scale of these disturbances, which were then thought to be driven by the presence of the inner edge, are on the order of the maximum "throw" distance of ejected material (~ $10^2$ - $10^3$ km), with the direction of propagation tending to be in the opposite sense to the ejecta distribution (i.e., inward for prograde), and are primarily governed by angular momentum exchanges. These kind of undulatory structures not associated with known resonances are observed in the C Ring, and throughout the B Ring (e.g., see Horn and Cuzzi 1996; Estrada and Cuzzi 1996). Durisen (1995) used a local linear instability analysis to demonstrate that ballistic transport causes these undulatory structures to arise spontaneously in an otherwise uniform ring system due to a linear instability, and did not require driving by the presence of an inner edge. Durisen concluded that the ~100 km structures seen in the inner B Ring (and possibly the more suppressed undulations seen in the C Ring) arise as a result of ballistic transport process, and perhaps viscous overstability, as suggested by some authors on the basis of stellar occultations (Colwell et al. 2007). However, in order to maintain the sharp inner edge, and produce the inner B Ring structure, a low-speed ejecta distribution component (that characterizes disruptive-type impacts) may be required in addition to the prograde distribution.

Finally, Durisen et al. (1996) considered the "direct" effects of meteoroid bombardment on the rings which include mass deposition (of the incoming projectiles), radial drifts that arise due to mass loading and/or loss of ejecta angular momentum, and radial drifts due to torques caused by the asymmetric absorption of the aberrated meteoroid flux (e.g., Cuzzi and Durisen 1990). The main result these authors find is that the effects of radial drifts due to mass loading are dominant by an order of magnitude, and that its most notable effect would be to cause parts of the Saturn ring system (i.e., the C Ring) to be lost to the planet in a timescale less than the age of the Solar System ($10^7$ to $10^8$ years, assuming the flux of Cuzzi and Estrada 1998) due to a decrease in the ring material's specific angular momentum. The result, however, is subject to uncertainties in the rate of mass deposition. Conversely, if the rings are old, there would have been significant mass loss over the age of the Solar System, and we would need the C Ring to somehow be "regenerated" regularly, possibly from spillover of material from the B Ring. Nonetheless, it would still only require some small fraction of the age of the Solar System for the C Ring to darken to its current state from the B Ring color (whereas the B Ring color presumably represents the effects of meteoroid bombardment over the age of the Solar System, see below). Then the C Ring we see now might be some grandchild of an "original C Ring", with the timescale for recycling of the C Ring likely being on the same order as the C Ring lifetimes discussed earlier. Unfortunately, such a scenario has yet to be studied in detail.

### 3.1.3 Spectral evolution

Subsequently, Cuzzi and Estrada (1998) developed a "pollution transport" code to model the evolution of ring composition with time under the influence of meteoroid bombardment and ballistic transport. Their goal was to see whether the meteoroid bombardment process was consistent with the regional and local color differences in Saturn's ring system given a uniform "primordial" or intrinsic composition, and if so, to attempt to constrain the intrinsic and extrinsic (bombarding) materials, and provide an independent estimate of ring age. The need for both intrinsic and extrinsic sources of material stems from the argument that if the rings started as pure water ice, then meteoritic bombardment by dark, spectrally neutral material such as carbon black (or even something slightly red and dark) would explain the C Ring color fairly well, but one could never produce the B Ring color. On the other hand, if the incoming material were spectrally red in color, then one could drive the B Ring to its present observed color, but the C Ring would be even more red than the B ring. The assumption, then, was that the red material was primarily intrinsic in origin (which might be consistent, e.g., with the breakup up of a small icy satellite that formed in the Saturnian subnebula, see Sec. 4.1) and that the meteoritic material bombarding the rings was dark and neutral in color. Their model consisted of two parts. The first was a dynamical code developed from the structural code of Durisen et al. (1989, 1992) that assumed time-invariant structure (e.g., constant optical depth, surface density), in which they calculated how the fractional amounts of non-icy constituents build up over time and how these impurities are redistributed over the rings. These authors assumed time-invariant structure because Durisen and colleagues had found in their structural evolution studies that constant optical depth regions and inner edges can remain more or less unchanged for very long timescales. The second part was a radiative transfer calculation that used the results of the ballistic transport code to see how ring particle composition was manifested in ring particle color and albedo.

Using their radiative transfer code, Cuzzi and Estrada (1998) calculated the albedo of a ring particle using reflectance values obtained from three different Voyager filters (G, V, and UV, which corresponded to 562, 413, and 345 nm, respectively; Estrada and Cuzzi 1996; also see Estrada et al., 2003) by volumetrically mixing together icy and non-icy constituents with different mass fractions and real and imaginary indices of refraction. An implicit assumption made by these workers was that in the low-phase-angle Voyager geometry, ring brightness was dominated by single scattering, and that the particle phase function was independent of wavelength so that ring reflectivity (color) ratios were just ratios of ring particle albedo at two wavelengths (however, see below). This assumption allowed them to calculate the ring color at two extreme radial locations (which bracketed the B Ring/C Ring transition) in order to determine the "current" refractive indices of ring material at these locations. The two reference points were treated as identical initial "primordial" composition, which evolved independently to their current composition due to the deposition of meteoritic material. Primordial (i.e., initial) composition was then inferred using a linear volume mixing approximation, while the composition of the extrinsic material was determined through iteration using the results of the ballistic transport code for the fractional amounts of non-icy constituents.

The final stage of the modeling process involved fitting all three radial color profiles (G/UV, G/V, V/UV) using their mixing model (Fig. 6). It should be noted that the mass mixing ratios of the individual constituents in their simple two-point model can only provide a constraint on the absolute ring age if the micrometeoroid flux, retention efficiency, and impact yields are known. They resolved this ambiguity by modeling the actual shape of the B Ring/C Ring transition, because it turns out that the shape depends on a combination of all these parameters. The best fits of Fig. 6 correspond to an estimated timescale of ~ 3-4 $\times 10^8$ years, much less than the age of the Solar System. Thus, Cuzzi and Estrada (1998) found that they could simultaneously explain both the C Ring/Cassini Division versus A Ring/B Ring dichotomy and the form and shape of the B Ring/C Ring transition on a timescale similar to that found by Durisen and colleagues. Moreover, the structural evolution conclusions and timescales and the conclusions of the pollution models were obtained independently of one another, making the overall result quite robust.

The actual age of the rings derived from these models remains quite uncertain because the absolute timescale for ring erosion depends on the typical impact yields of ejecta; the efficiency with which extrinsic material retains its absorptive properties (rather than changing drastically in composition during high-speed impacts onto the rings); and the micrometeoroid flux. It should be noted that, although the historical scenario is that the rings are primordial (e.g., Pollack 1976), and the post-Voyager paradigm became that they are young, as described above (also see section 2), some workers have revived the idea that the rings could be as old as the Solar System. For example, Esposito et al. (2005) find variations in ring brightness over scales of 1000-3000 km which they assert are not clearly consistent with ballistic transport, and posit that a local event such as the breakup of a small icy moon may be responsible for resetting the "ballistic transport" clock. Stewart et al. (2007) and Robbins et al.(2009) suggest that gravitational instabilities in massive rings could easily hide most of the mass in dense, but transient, clumps and streamers, while a small fractional percentage of fairly empty gaps between these opaque structures would then provide the observed ring transparency or finite optical depth. Esposito (2006) suggested that if the ring surface mass density is an order of magnitude larger than currently inferred from optical depth and sparse spiral density wave measurements, the rings can more easily resist becoming polluted and darkened, and could perhaps indeed be as old as the Solar System (see Sec 3.2.1). A caveat to keep in mind, however, is that the micrometeorite flux was certainly higher in the early Solar System. As much as an order of magnitude (or more) higher is not out of the question, which might require that an ancient B Ring have even more mass than what is currently theorized to remain as icy as observed, despite the cumulative effects of pollution over such a long period of time.

In the midst of the Cassini era, a renewed vigor has emerged regarding ballistic transport modeling. Dozens of occultations cutting all parts of the main rings at a number of longitudes and elevations is allowing for a more detailed understanding of ring properties such as opacity, surface density, optical depth, and ring viscosity – all of which are key quantities for structural evolution modeling. On the other hand, a wide range of Cassini spectral data which spans a larger range of the electromagnetic spectrum in many combinations and viewing geometries than previously obtained by Voyager is allowing for a comprehensive compositional study. For example, the assumption made by Cuzzi and Estrada (1998) that ring particle phase functions are wavelength independent turns out to be incorrect (Cuzzi et al. 2002; Poulet et al. 2002; see chapter by Cuzzi et al.). Thus modeling ring color and composition requires a more sophisticated approach which involves a combination of ring layer and regolith radiative transfer modeling. Fortunately, given the wide array of Cassini spectral data available, the problem is a tractable one.

## 3.2 Limited accretion

As seen in the previous section, there are several processes capable of eroding and altering the bulk composition of the ring system, and thus a renewing mechanism is often invoked (see sections 2 and 3.4). A combination of accretion and erosion could be such a mechanism. Here we present how a limited form of accretion could take place inside the rings, despite the tidal forces. Further discussion of accretion physics can be found in the chapter by Schmidt et al.

The concept of accretion is embodied in the convenient, but sometime confusing, notion of the "Roche Limit", which depends on several assumptions (fluid vs. rigid body, etc.). Depending on these assumptions, the Roche Limit is located somewhere between 1.26 R (where R is the planet's radius, and assuming the body has the same density as the planet, section 2.2.1) and 2.456 R, corresponding, respectively, to the inner C Ring and about 8,000 km beyond the F Ring. So it is plausible that substantial accretion could occur in Saturn's rings; however, it should be severely limited by Saturn's strong tides. Indeed, the Cassini spacecraft reveals indirect evidence for accretion. The main observations are the following:

(1) The detection of self-gravity "wakes" in Saturn's A Ring (Colwell et al., 2006; Hedman et al., 2007b) thanks to multiple stellar occultations. Wakes are a gathering of particles due to a local gravitational instability (see chapter by Schmidt et al. and below).
(2) The detection of propeller-shaped structures in Saturn's A Ring (Tiscareno et al., 2006, 2008; Sremčević et al., 2007), revealing the presence of chunks of material in the 50-100 m range. Propellers could be either the result of fragmentation of an ancient moon (Sremčević et al., 2007), or, the result of "creeping" accretion processes (Esposito et al., 2008b).
(3) VIMS data have revealed that the regolith on the surface of Saturn's ring particles shows a general trend for increasing grain size with distance across the rings (Nicholson et al., 2008).
(4) ISS images of Pan and Atlas, two small satellites embedded in, or close to, the rings have revealed prominent equatorial bulges on these moons (Porco et al., 2007). Simulations show that these bulges are likely the results of ring particles accreted at the satellites' equators (Charnoz et al., 2007)
(5) ISS surveys of the F Ring have revealed the presence of embedded bodies that could not be tracked for more than a couple of orbits (Porco et al., 2005) suggesting the idea of "ephemeral moonlets". Such bodies might be produced by accretion. Stellar occultations of Saturn's F Ring (Esposito et al., 2008a) have revealed the presence of clumps embedded in the ring's core that could well be the result of some accretion process owing to their optical translucency, suggesting loose aggregates of material (Fig. 2).

All this indirect evidence supports the idea that accretion may indeed take place in Saturn's rings, and that an active recycling of material is possible. In the last twenty years several analytical studies and numerical models have characterized accretion processes under tidal forces (e.g., Weidenschilling et al., 1984a, Ohtsuki 1993, Canup and Esposito 1995, Barbara and Esposito 2002, Karjalainen and Salo 2004, Albers and Spahn 2006), and have shown that an exotic form of accretion may take place in the rings.

### 3.2.1 Gravitational instability

A gravitational instability occurs when the local self-gravitational potential energy exceeds both internal energy (due to pressure) and shear kinetic energy (due to keplerian shear, i.e., the variation of the ring particles' angular speed with distance from the planet (Eq. 4)). It is embodied in the "Jeans-Toomre" criterion, implying that the gravitational instability occurs for small Q, such that (Toomre 1964, Karjalainen and Salo 2004, and see chapter by Schmidt et al.):

$$Q = \frac{c\Omega}{3.36 G \sigma} \qquad \textbf{Eq. 10}$$

where c is the velocity dispersion, G is the gravitational constant, $\Omega$ is the local keplerian angular velocity and $\sigma$ is the surface density. For Q < 2 the collective gravity together with keplerian shear creates shearing, tilted wake structures. Numerical simulations (Salo 1995; Karjalainen and Salo 2004) show that gravitational wakes are progenitors of gravitational aggregates: in the inner regions of the A Ring, wakes are like parallel rods with moderate density contrast with the inter-wake medium. Further out in the A Ring, the density contrast increases and the wakes become more and more clumpy. Finally, just outside of the A Ring, the wakes coalesce into clumps, recalling small satellites (see, e.g., fig. 2 of Karjalainen and Salo 2004). For more details on wakes, the reader is invited to read the chapters by Schmidt et al. and Colwell et al.

### 3.2.2 Tidally modified accretion

Even in the absence of gravitational instability (i.e., when Q > 2), accretion may still be possible via binary encounters, as for planetesimal growth in the protoplanetary disk. The classic two-body criterion for accretion is simple: the rebound velocity, $V_r$ (which is a fraction $\varepsilon < 1$ of the impact velocity $V_i$), must be less than the escape velocity of the two bodies. This is valid in free space, but becomes incorrect under the tidal influence of a nearby massive body (Saturn in our case), called the "primary": close to the primary, the differential tidal acceleration can overcome the two bodies' gravitational acceleration. To describe 3-body modified accretion, it is necessary to properly take into account the potential energy of the colliding bodies. We follow the approaches of Ohtsuki (1993) and Canup and Esposito (1995). Consider a body with mass m and radius a, orbiting on a circular orbit at distance r from Saturn, with keplerian angular velocity $\Omega$. The specific potential energy of a body with mass m', radius a', located at r', in the (non-inertial) frame rotating with body m is:

$$E_p = \frac{-GM_s}{r'} - \frac{-Gm}{\|\vec{r} - \vec{r}'\|} - \frac{\Omega r^2}{2} \qquad \textbf{Eq. 11}$$

After developing Eq. 11 to first order in (r-r') and nondimensionalizing time scales with $\Omega^{-1}$ and lengths with the Hill radius ($R_{hill}=r\ [(m+m')/(3M_s)]^{1/3}$), and using relative coordinates in the rotating frame x,y,z (Nakazawa and Ida 1988; Ohtsuki 1993) so that $r=(x^2+y^2+z^2)^{1/2}$, we get (see section "Accretion of particles in the Roche Zone" in the chapter by Schmidt et al. for more details):

$$E_p = -\frac{3}{2}x^2 + \frac{1}{2}z^2 - \frac{3}{r} - \frac{9}{2} \qquad \textbf{Eq. 12}$$

We continue, following the simplified model of Canup and Esposito (1995), which preserves the essential physics well enough for our purposes. When the two bodies collide, the scaled rebound velocity is computed as (after averaging over all impact angles) $V_r=\varepsilon[V_b^2+V_e^2+2/3r_p^2]^{1/2}$, with $r_p=(a+a')/R_{hill}$. In this expression $V_b$ is the relative velocity (the quadratic sum of the random velocity plus the gradient of orbital velocity due to different semi-major axes), $V_e$ is the scaled mutual escape velocity (= $(6/r_p)^{1/2}$), and $2/3\, r_p^2$ stands for the finite keplerian velocity difference between the two mass centers (see Canup and Esposito (1995) for details). Finally, the requirement for gravitational sticking is that the total energy $E= \frac{1}{2}V_i^2+E_P$ must be negative after rebound. In the limit of zero random velocities among particles (i.e., when relative velocities are simply due to keplerian shear), new behaviors appear. Since the potential energy cannot be arbitrarily negative (due to the finite radii of both bodies setting a lower bound to their distance) it turns out that even for $\varepsilon=0$ (perfect sticking in free space), no sticking occurs for particles above a given size, so two large bodies are always separated from each other by the tidal force. More precisely, the condition $E<0$ defines a critical coefficient of restitution $\varepsilon_{crit}$, above which accretion is impossible, in the case of zero random velocity:

$$\varepsilon_{crit} = \left[ \frac{v_e^2 + 2/3 r_p^2 - 9}{v_e^2 + v_b^2 + 2/3 r_p^2} \right]^{1/2} \quad \textbf{Eq. 13}$$

$\varepsilon_{crit}$ is displayed in Fig. 7, showing a steep fall to 0 for $r_p \geq 0.691$ (the condition for zero numerator in Eq. 13, see Canup and Esposito 1995). Note that the case of non-zero random velocity has been treated in Ohtsuki (1993), and it is found that accretion is still be possible in this case, although severely modified by tides, provided the coefficient of restitution is low enough (see the chapter by Schmidt et al. for details).

The sticking requirement $r_p=(a+a')/R_{hill}\leq 0.691$ means that the sum of the physical radii of the particles must be smaller than the Hill sphere of the agglomerate, i.e., the two bodies' mass centers must be gravitationally bound, despite the tidal forces, in order to stick. This requirement prevents accretion between equal size bodies, but allows accretion between unequal size bodies (a small particle at the surface of a big one); indeed, a small particle can be stored in the empty spaces of the Hill sphere of the big particle. In consequence, we would expect gravitational aggregates to naturally adopt the shape of Hill spheres. Another way of understanding this result is that for growth to occur, an aggregate must be denser than the average density of its own Hill Sphere ($\rho_{HS}$), given by (assuming a point mass, see Porco et al., 2007):

$$\rho_{HS} \approx \frac{3M_s}{1.59 r^3} \quad \textbf{Eq. 14}$$

Finally, tidally modified accretion implies a critical mass ratio, so that $m/m'$ must be larger than some factor, about 100 to 1000 in the A Ring (Canup and Esposito 1995). In the limit of zero mass ratio, and under the assumption of zero random velocity, this also yields a new definition of the Roche Limit $\alpha$:

$$\frac{\alpha}{R} < 2.09 \left( \frac{\rho}{\rho_p} \right)^{-1/3} \quad \textbf{Eq.15}$$

Eq. 15 sets the Roche Limit for ice at ~ 138,000 km, in very good agreement with the A Ring's outer edge at ~136,800 km. Numerical simulations (Karjalainen and Salo 2004; Karjalainen 2007) have qualitatively confirmed all the previous considerations. Further discussion can be found in the chapter by Schmidt et al.

### 3.2.3 Surface sticking

Up to now, only gravitational accumulation was considered. As a direct consequence, a gravitational aggregate of any size can grow inside the rings as long as its density remains high enough so that the body is geometrically contained inside its Hill sphere: numerical simulations shows that it should grow without bound, as long as it is immersed in the disk (Porco et al., 2007). However, for a body with internal strength (for example, due to intrinsic material strength, surface sticking, etc.) there should be a size vs. distance dependence because the tidal stress may exceed the material strength at some locations. The taking into account of material strength was done in Davis et al. (1984) and in Weidenschilling et al. (1984a) and introduced the notion of "Dynamical Ephemeral Bodies", i.e., bodies that grow to a critical size, determined by the balance between their internal strength (gravitation + material strength) and tidal stress, before being torn apart when the tidal forces exceed their internal strength. A recent work (Albers and Spahn 2006) introduced an efficient formalism to take into account adhesive forces, using a viscoelastic model of collisions. When two particles stick on each other, their contact surface is (to first order) a circle with radius a, requiring a surface energy $U_s$ given by :

$$Us = -\pi \gamma a^2 \qquad \textbf{Eq. 16}$$

where $\gamma$ is the surface energy per surface unit. Eq. 16 shows that larger contact areas will result in higher surface energy. Solving the collision dynamics shows that above a certain size, smaller grains favorably stick to larger ones. This translates into an adhesion force (Albers and Spahn 2006):

$$F_{ad} = 2/3 \gamma \pi R_{eff} \qquad \textbf{Eq. 17}$$

where $R_{eff}=R_1 R_2/(R_1+R_2)$ is the effective radius of particles with radii $R_1$ and $R_2$. The value of $\gamma$ is poorly known, but is estimated to be about 0.74 N/m for ice (Albers and Spahn 2006). By setting $F_{ad}$ plus the mutual gravitational force equal to the tidal acceleration, a new criterion is established for the stability of a two-particle aggregate with a dependence on the particle size. Let $\mu=R_2/R_1$ and $R=R_1$. A new "Roche Limit" ($R_{crit}$) can be defined as the distance where a stable aggregate can exist (eq. 18 of Albers & Spahn 2006):

$$R_{crit}^3 = \frac{24GM_s \rho R^3 \mu^2 (1+\mu)^3 [(k+1)^2+2]}{(1+\mu^3)[27\gamma(1+\mu)+32\pi G \rho^2 \mu^2 R^3]} \qquad \textbf{Eq. 18}$$

where k is the order of 1 and is proportional to the rotation frequency of the particle. We now see that (i) a dependence on the particle size has appeared and (ii) that $R_{crit}$ also depends on the size ratio $\mu$. $R_{crit}$ is plotted for R=10 m in Fig. 8. We see that for equal size bodies ($\mu$=1), agglomerates up to R = 10 m can be stable in the D Ring and up to R = 100 m for the B Ring. Conversely, in the D Ring, aggregates with R > 10 m are stable only for $\mu$ much larger than 1, corresponding to aggregates of smaller grains sitting on the top of much larger ones.

### 3.2.4 Accretion of small embedded satellites?

At this point a natural question is: did the small satellites embedded within or just outside Saturn's main rings (e.g. Pan, Daphnis, Atlas) accrete *inside* Saturn's rings? Recent Cassini observations have provided high resolution images and have allowed a determination of the shapes of Pan and Atlas (Porco et al. 2007). They are significantly flatter than Hill spheres, thus suggesting they are not the result of pure tidally modified accretion inside the rings (Charnoz et al., 2007; Karjalainen 2007; Porco et al., 2007). Pan and Atlas exhibit prominent equatorial ridges whose locations are well reproduced in simulations of ring particle accretion onto the surface of pre-existing bodies (Charnoz et al., 2007). This suggests that both Pan and Atlas may hide an inner core, on top of which an envelope of ring particles was accreted. Computation of the gravitational potential at their surface shows that the equatorial ridges extend beyond their Hill apheres (Charnoz et al., 2007), implying that the ridges are maintained by inter-particle adhesion forces, or material strength, rather than pure gravitation.

In conclusion, there are numerous reasons to think that there is ongoing accretion in Saturn's rings; however, its characteristic seems very sensitive to the particles' size, shape and location, as well as the physics at work (gravitation, surface sticking, etc.). There should be a general trend for bigger bodies as a function of distance from the planet. Analytical models are supported, at least qualitatively, by numerical simulations. More extensive accretion may occur in the F Ring region, which in turn, may give us clues to understanding some of the peculiar aspects of this ring (section 5).

### 3.3 Collisional cascade

We have seen in the previous section that under some circumstances, some limited accretion could be possible in the rings, especially near their outer edge. In order for material to recycle back into the rings, a destruction process is needed. Destruction of the biggest bodies, or moonlets, via meteoroid bombardment could lead to a cascade of smaller collisions. A moon shattered by a large impact from an interplanetary projectile would become a ring of material orbiting the planet. Big moons are the source of small moons; small moons are the source for rings. Rings are eventually ground to dust that is lost by becoming charged and carried away by the planet's rotating magnetic field or by atmospheric drag into the planetary atmosphere (where it shines briefly as a meteor), and maybe produce a "ring of fire" as suggested by Rubincam (2006). This process is called a "collisional cascade" (see Fig. 9).

Since the shattering of a moon is a random event, ring history will be stochastic and somewhat unpredictable. The differences between the various ring systems might be explained by the different random outcomes of this stochastic process. Thus, the collisional cascade can provide an explanation for the apparently different ring systems around each of the giant planets. Catastrophic events provide the tempo for creating planetary rings: new rings are episodically created by destruction of small moons near the planet (Colwell et al., 1992, 2000). This disorderly history arises from singular events.

The most serious problem with this explanation is that the collisional cascade uses the raw material (a planet's initial complement of moons) too rapidly. If we imagine we are now looking at the remnants of 4.5 billion years of successive destruction over the age of the Solar System, then this process is almost at its end. The small moons that now remain as the source of future rings have a lifetime of only some few hundred million years, based on calculations by Colwell et al. (2000). This is less than 10% of the age of the Solar System. Why are we humans so fortunate as to come upon the scene with robotic space exploration, just in time to see the rings' finale? A direct consequence of this process for Saturn's rings is the destruction of the population of moonlets near the outer edge of the rings, like Pan, Daphnis and Atlas.

### 3.4  Ring-moon interactions

Whereas ring-moon interactions do not seem directly connected to the question of ring origin and evolution, dynamical interactions between rings and satellites can result in rapid orbital evolution; indeed, this rapid evolution provides the original argument for the possible youth of Saturn's rings. Unfortunately the torque exerted by the ring on a moon is complex to compute and depends on several assumptions and on the ring surface density, which is poorly known. This topic is discussed in detail in the chapter by Schmidt *et al.*, but we briefly outline the relevant physics here.

Consider a ring particle and a satellite orbiting exterior to it, with both bodies on low-eccentricity, low-inclination orbits around Saturn. At an arbitrary location in the rings, the frequency at which the satellite perturbs the ring particle will not be a natural oscillation frequency of the particle (such as its radial (epicyclic) frequency, $\kappa$), and the satellite will have no systematic effect on the ring particle's orbit. However, at the satellite's inner Lindblad resonances (ILRs), a more dramatic interaction can take place. The location of these resonances is given by the condition $r = r_L$, where

$$\omega = m\Omega(r_L) - \kappa(r_L). \qquad \textbf{Eq.19}$$

In this expression, $\omega$ is the forcing frequency due to the moon; m is a positive integer; $\Omega(r_L)$ is the orbital frequency of a ring particle at $r_L$; and $\kappa(r_L)$ is the radial frequency of the ring particle (Shu 1984). In the simplest case, $\omega = m\Omega_M$, where $\Omega_M$ is the moon's orbital frequency. If Saturn were a sphere, its potential would be keplerian, like a point mass, and the orbital and radial frequencies of a ring particle would be equal. Substituting $\omega = m\Omega_M$ and $\Omega(r_L) = \kappa(r_L)$ in Eq. 19, we find

$$m\Omega_M = (m - 1)\,\Omega(r_L), \qquad \textbf{Eq. 20}$$

which is known as an m:(m-1) resonance because a ring particle completes m orbits for every m – 1 orbits of the satellite. For example, the outer edges of the B Ring and A Ring lie near the 2:1 and 7:6 resonances with Mimas and Janus/Epimetheus, respectively. The Prometheus 2:1 ILR lies in the C Ring, while the Prometheus 6:5, 7:6, ..., 33:32, and 34:33 ILRs lie in the A Ring (Lissauer and Cuzzi 1982; Nicholson et al. 1990; Tiscareno et al. 2007).

By Kepler's third law, $r_L = r_M[(m-1)/m]^{2/3}$, where $r_M$ is the semi-major axis of the moon's orbit. For large m, successive resonances are separated by $2r_M/(3m^2)$, which is less than 100 km for the ILRs of Prometheus and Atlas in the outermost part of the A Ring. (In reality, Saturn's oblateness causes $\kappa$ to be slightly smaller than $\Omega$, causing these resonances to fall slightly further from Saturn than in the keplerian case. In general, this shift does not affect the resonance dynamics or relative locations in any important way.)

At the inner Lindblad resonances, there is a torque which transfers angular momentum outward from the ring to the satellite. In the main rings, particularly in the A Ring, the rings' response to the satellite takes the form of a density wave that propagates outward from the resonance. The density wave is a nonaxisymmetric pattern which exerts a backreaction on the forcing satellite, causing it to recede from the ring. Goldreich and Tremaine (1980, 1982) calculated the torque using a linear theory; but even for strong (nonlinear) waves, the torque appears to be similar to the value given by linear theory (Shu et al., 1985). In the case of a satellite of mass $m_s$ at distance $r_M$ from the center of Saturn that is close to a ring of surface mass density σ at distance r ("close" means that $r_M - r \ll r$), the torque is proportional to $m_s^2 \sigma (r_M - r)^{-3}$, and the timescale for the satellite's orbit to expand is proportional to $(r_M - r)^3/(m_s\sigma)$. Assuming that they started at the outer edge of the A Ring, Prometheus and Atlas would have taken only some 10-100 million years to reach their current positions. (These timescales are about 10 times larger than those given by Goldreich and Tremaine (1982), because the satellites are less massive and the rings' surface density in the outer A Ring is much smaller than they assumed.) Some tens of millions of years in the future, Prometheus will cross the orbit of Pandora, and the satellites are likely to collide with each other (Poulet and Sicardy 2001). We may wonder what wouold be the result of such a collision: will they merge into a single new satellite one or will they be destroyed ? Since they are located close to the Roche Limit for ice (see section 3.2) a variety of outcomes are possible that will mainly depend on their impact velocity. It would not be impossible that a dusty ring form after the impact, but we have no idea if it will similar to the actual F ring or not.

When the Cassini mission was being planned, it was expected that the expansion of the orbits of some of the ring moons due to density wave torques would be directly measurable as a lag in their longitudes that would increase over the course of the mission. However, the dynamics of the ring moons has proven to be quite chaotic; for example, overlapping 121:118 resonances between Prometheus and Pandora cause "jumps" in their orbits at ~6-year intervals (e.g., French et al., 2003; Goldreich and Rappaport 2003a, 2003b ; Renner et al. 2005; Farmer and Goldreich 2006; Shevchenko 2008). As a result, the predicted recession of the moons from the rings have not yet been detected. In principle, the ring moons' orbital evolution could be much slower if the moons were resonantly locked to larger satellites (Borderies et al., 1984), but no such link is known to exist.

Satellites produce many other dynamical effects on rings. These include the formation of gaps at strong isolated resonances, such as the inner edge of the Cassini Division, which is associated with the Mimas 2:1 ILR; gaps due to the overlapping resonances of embedded satellites, such as Pan and Daphnis, which produce the Encke and Keeler Gaps, respectively; "shepherding," or confinement, of narrow ringlets; and the excitation of the random velocities of ring particles in wave regions due to the energy input by the wave. This dynamical "heating" increases the viscosity, and hence spreading rate, of the ring (see section 2.2.2). For details, we refer the reader to the reviews by Cuzzi et al. (1984), Esposito et al. (1984), Lissauer and Cuzzi (1985), and the chapter by Schmidt et al. in this volume.

## 3.5 The long term evolution of Saturn's main rings
From the different processes described above, is it possible to draw a picture of the long term evolution of Saturn's rings? This is a difficult task since the interactions between these effects have never been studied. In addition local effects (like accretion) and large scale effects (like meteoroid bombardment) are difficult to couple in the same formalism or simulation. We now try to briefly imagine what the long term evolution of the rings might be like. First, viscous evolution and meteoroid bombardment would likely spread the C Ring (and perhaps the B ring itself) closer to the planet, on timescales less than $10^9$ years. The C Ring could also be replaced by material leaking from the B Ring (through meteoroid bombardment, for example). Small moonlets orbiting at the edge of the rings will migrate to larger radii, and be trapped in mutual resonances, perhaps into horseshoe or tadpole orbits (like Janus and Epimetheus) on timescales of about $10^8$ years. If the resonant trapping is ineffective, collision and re-accretion could occur, creating a new generation of bigger satellites from the former generation of smaller ones. Whereas the torque exerted by the rings on Mimas and Janus/Epimetheus are not well determined because their strongest resonances lie near the outer edges of the B and A Rings, respectively, these satellites should migrate to larger radii. It is even possible to imagine a distant future in which the Cassini Division has shifted outward due to the outward movement of the 2:1 Mimas resonance (which establishes the inner edge of the Cassini Division). In addition, since Mimas is in a 4:2 mean motion resonance with Tethys, the ring torque should be transferred simultaneously to both moons. The A Ring is expected to viscously spread out beyond the Roche limit in $10^8$ years (due to the recession of Janus/Epimetheus), and new moonlets should form at this location due to the sharp fall of tidal forces at the Roche Limit. The final fate of the rings is very difficult to imagine, at least because we do not really understand how the rings could ever exist today. Viscous spreading and meteoroid bombardment could be the ultimate mechanisms, acting without end until the rings have finally fallen into the planet and spread outside the Roche Limit. However, so many mechanisms remain poorly understood that these conclusions should be taken with much care.

## 4. Scenarios for origin of the main rings

The origin of the faint rings of the giant planets seems to be quite well understood: with the exception of Saturn's E Ring, they are attributed to the erosion and periodic destruction of km-sized moonlets due to meteoroid bombardment (see section 6). However, due to their very high mass and putative young age, such scenarios cannot apply directly to Saturn's massive ring system. Keeping in mind that we still do not understand how the rings could have formed in the last billion years, we try here to present the details of three scenarios for the origin of Saturn's main rings, and present the pros and the cons for each of them. The first scenario (section 4.1) suggests that Saturn's rings are a remnant from Saturn's primordial nebula, the second one suggests that Saturn's rings are debris from a destroyed satellite (section 4.2), and the last one suggests that they are debris of one or several comets tidally disrupted by Saturn (section 4.3).

### 4.1 Remnant from Saturn's sub-nebula disk?

The circumplanetary gas disk or subnebula is a by-product of the later stages of giant planet accretion. Its formation occurs over a period in which the giant planet transitions from runaway gas accretion to its eventual isolation from the solar nebula, at which time planetary accretion ends. This isolation occurs because the giant planet has either tidally opened a well-formed gap, effectively pushing the nebula gas away from itself, or the nebula gas dissipates. The formation of the regular satellites is expected to occur towards the tail end of giant planet accretion when the planet is approaching its final mass, and at a time when any remaining inflow of nebula gas through the giant planet gap is weak and waning. A more detailed discussion of satellite accretion in the context of the combined process of giant planet and circumplanetary disk formation may be found in Estrada et al. (2008; see also chapter by Johnson et al.). If the rings are the remnants of Saturn's subnebula, then their formation must have taken place in the same environment in which the satellites formed, which must be discussed first.

#### 4.1.1 Satellite formation
Satellite formation appears to be a natural consequence of giant planet accretion. The fact that all of the giant planets have ring systems suggests that their origin results from processes common to the giant planets' environments. The observed giant planet satellite systems are quite diverse in appearance and composition, and the rings are no less so. The Jovian rings are ethereal, diffuse, and composed of non-icy material; those of Uranus and Neptune appear to be "dusty", dark and neutral in color (e.g. Cuzzi 1985; Baines et al., 1998; Karkoschka 2001), although some Uranian rings have been found to be spectrally red or blue (e.g., de Pater et al., 2006). On the other hand, the rings of Saturn are prominent and massive, composed mostly of water ice, but distinctly reddish in color, suggesting the presence of organic material (e.g., Estrada and Cuzzi 1996; Cuzzi et al., 2002; Estrada et al., 2003), possibly tholins, in addition to some neutrally colored darkening agent (Cuzzi and Estrada 1998; Poulet et al., 2003). The darkening of the rings over time can most likely be associated with extrinsic meteoroid bombardment (see Sec. 3.3.1). However, if the rings of the giant planets did form during the time of satellite formation, then their different compositions might, in great part, reflect differences in their respective accretional environments. It would also imply that at least some parts of the rings are as old as the Solar System, a conclusion that may be at odds with the various lines of evidence that suggest that the rings are geologically young.

**4.1.2 Implanting the ring system**
The model in which the rings formed in a relatively massive subnebula from essentially the same processes that lead to the formation of the satellites has been referred to as the "condensation" model (Pollack 1975; Pollack et al., 1977; Pollack and Consolmagno 1984). In this model the ring particles form through the sticking of sub-micron sized dust grains which are dynamically coupled to the subnebula gas, and collide at low (size-dependent) relative velocities that can be caused by a variety of mechanisms (e.g., Völk et al., 1980; Weidenschilling 1984b; Nakagawa et al., 1986; Ossenkopf 1993; Ormel and Cuzzi 2007). As grains grow into agglomerates, they begin to settle to the midplane, and may continue to grow through coagulation and coalescence, and as the subnebula cools, through vapor phase deposition. In the satellite forming regions, particles may continue to grow further as they settle by sweeping up dust and rubble (Cuzzi et al., 1993; Mosqueira and Estrada 2003a), eventually growing large enough that they may "decouple" from the gas; however, close to the planet, specifically within the Roche zone where tidal forces begin to overcome gravitational sticking, the final stages of growth are stymied, so that one tends to be left with a population of smaller particles (Pollack and Consolmagno 1984). In addition, growth close to the planet, where dynamical times are quite short relative to the solar nebula, is likely further limited simply as a natural consequence of relative velocities between particles entrained in the gas being too high to allow coagulation beyond some fragmentation barrier. For example, the dynamical times at the A Ring would be comparable to those at ~ 0.01 AU in the solar nebula (a distance at which refractories may not condense due to high temperatures in any case). In spite of these difficulties, the main issue facing this picture is how these small particles can survive long enough for the subnebula gas to dissipate.

In the condensation model, the composition of the rings being primarily icy has been attributed to two things, both which lead to the formation of ice particles late in the lifetime of the subnebula. The first is that the atmosphere of Saturn, even after envelope contraction, extended beyond the region of the rings. In the early stages of gas accretion long before satellite formation can even begin, the giant planet's atmosphere fills up a fair fraction of its Hill sphere (Bodenheimer and Pollack 1986; Pollack et al., 1996); but, once envelope contraction happens it occurs fairly rapidly compared to the runaway gas accretion epoch, which lasts ~ $10^4 - 10^5$ years (Hubickyj et al., 2005), and at a time when the planet is only a fraction of its final mass (Lissauer et al., 2009). The exact timing of the collapse depends on several factors (see Lissauer and Stevenson 2007).

### 4.1.3 Collapse and cooling of the envelope

The envelope collapse results notably in two things relevant to rings and satellites: the formation of a compact subnebula disk component due to the excess angular momentum of accreted nebula gas in the envelope prior to collapse (Stevenson et al., 1986; Mosqueira and Estrada 2003; Estrada et al., 2009); and, as implied above, the planet's radius shrinks down to a few planetary radii (e.g., Lissauer et al., 2009, and references therein). Although contraction of the giant planet's radius to its current size takes the remainder of the planet's lifetime, by the time the subnebula gas dissipates, the planet's atmospheric boundary (within which temperatures at this time might still remain too high for condensation of water ice) may lie just within or interior to the radial location of the B Ring (e.g., Pollack et al., 1977). If so, the atmospheric boundary might provide a natural mechanism for explaining an initial sharp edge for the B Ring (which is subsequently maintained by resonant and ballistic transport processes). As Saturn continued to contract to its present size, gas drag may have leeched off particles from the inner B Ring, leading to the first incarnation of the C Ring, and as suggested in Sec. 3, other processes may be able to continually reproduce the short-lived C Ring over the age of the Solar System.

Second, the luminosity of Saturn had to decrease sufficiently to allow for the condensation of ice in the innermost regions of the circumsaturnian gas disk. During the beginning stages of satellite formation, when the planet is approaching its final mass and any remaining gas inflow from the nebula wanes, the planet's excess luminosity alone is enough to produce inner disk temperatures that would be initially too high to allow for the condensation of water (and depending on ambient subnebula conditions, condensates more volatile than water, e.g., Pollack et al., 1976). As a result, silicates would likely condense first, and be lost to the planet due to gas drag forces and/or settling. In this way, the subnebula may be "enhanced" in water ice and depleted in rocky material, particularly in the ring and mid-sized satellites region (Pollack et al., 1976; Mosqueira and Estrada 2003a,b; also see the chapter by Johnson et al. for a discussion of water-ice enrichment). Because the cooling time for the Saturnian subnebula post-planet accretion ($\sim 10^5$ years, Pollack et al., 1977) may be significantly shorter than the circumsaturnian disk lifetime ($\sim 10^6$-$10^7$ years, e.g., by photoevaporation, Shu et al., 1993), this enhancement would suggest that late stage accretion would lead to ice-rich satellites and rings. Contrast this with the Jovian subnebula, in which the cooling time is an order of magnitude longer, and comparable to the disk dissipation time (Pollack and Reynolds 1974; Lunine and Stevenson 1982). In this case, water ice may never condense in the inner regions prior to disk dissipation. If the rings are formed as a result of condensation, then the Jovian ring composition would be consistent with this picture.

### 4.1.4 The role of turbulence

The satellites are expected to accrete in an environment in which turbulence, driven by the gas inflow, decays (Mosqueira and Estrada 2003a,b). This presumes that there is no source of intrinsic turbulence that could continue to drive subnebula evolution once the gas inflow from the solar nebula is cut off. This assumption is supported by both numerical (e.g., Hawley et al., 1999; Shen et al., 2006), and laboratory experiments (Ji et al., 2006) that question the ability of purely hydrodynamical turbulence to transport angular momentum efficiently in Keplerian disks. Yet, it may be possible that in a scenario in which turbulence decays, some mechanism capable of driving turbulence may persist very close to the planet (Mosqueira and Estrada 2003a), which could provide another reason why the condensation of water ice can be delayed or prevented altogether.

For example, turbulence due to a magneto-hydrodynamic instability (MRI, Balbus and Hawley 1991) may apply, assuming that a combination of density, ionization, and temperature conditions allows for it (e.g., Gammie 1996; McKee and Ostriker 2007), but may be rendered ineffective due to the constant production of dust as a result of the persistent fragmentation of particles. Another potential driving mechanism that may operate is convection (e.g., Lin and Papaloizou 1980). The continuous fragmentation of particles due to both high systematic and turbulence-induced relative velocities within the Roche zone may keep the dust opacity high, allowing for a strong vertical temperature gradient to be sustained. However, if convection drives turbulence, the angular momentum transport may be quite weak (e.g., Stone and Balbus 1996), and furthermore, directed inwards (e.g., Ryu and Goodman 1992; Cabot 1996). This means that a "cavity" may be created close to the planet, which would effectively terminate gas accretion onto Saturn. The consequences of such a cavity on satellite and ring formation have not been explored.

**4.1.5 Caveats**
The rings' being primarily water ice seems to suggest that if they condensed from the subnebula at their present location, then turbulence and the viscous heating associated with it, at least at the tail end of the lifetime of the subnebula, was absent even close to the planet. Moreover, if water ice were allowed to condense, ring particles would still need to be able to survive being lost to the planet by gas drag, which is responsible for clearing wide regions of the circumplanetary disk (the essentially empty regions between the satellites). This difficulty can be overcome by presuming that the condensation of icy ring particles continued up until disk dispersal (Pollack 1975) which may allow for conditions in which gas drag may not have had enough time to clear the remaining solid material before the subnebula dissipated.

Another difficulty is explaining the presence of the small moonlets embedded in the rings, such as Pan and Daphnis. Pollack and Consolmagno (1984) suggested that their growth may have been enabled at resonance locations. Recently, Charnoz et al. (2007) and Porco et al. (2007) have concluded that these moons may have been collisional shards initially one-third to one-half their present sizes that grew to their present-day sizes by accreting ring material. In order to facilitate growth, a core of sufficient density is required in order to be stable against tidal disruption. Their growth is then mostly limited by tidal truncation (i.e., they open gaps), rather than tidal shear. An alternative possibility is that such objects may have been satellitesimals that drifted in via gas drag and were left stranded as the disk dissipated, and also grew to their present sizes by the accumulation of ring material. Note that gas drag and type I migration may have also provided a means for bringing in larger objects such as embryos or small moons close in to the planet where they may have eventually been broken up (see section 4.2.1).

If the rings are much younger than the age of the Solar System, then their origin is almost assuredly due to some collisional process. However, if the rings (at least the B ring) turn out to be much more massive than previously thought, then it allows for the possibility that the rings are as old as the Solar System (although, it should be noted that by themselves, massive rings do not imply ancient rings), and their origin is open to both collisional and condensation model hypotheses. The spectral shape of the ring material being most similar to that of the Saturnian mid-sized satellites (Barucci et al., 2008; although cf. Buratti et al., 1999 and section 3.1) may tend to support the idea that the materials that compose the rings and satellites were processed under similar conditions (e.g., processed in the subnebula). Given that it may only take a very small fraction of non-icy intrinsic material (< 1% if volumetrically mixed; Grossman 1990; Cuzzi and Estrada 1998; Poulet et al., 2003) to give the rings their reddish color, their composition does not appear inconsistent with the condensation scenario. Thus, although physically plausible if the rings are as old as might be allowed by a much higher mass than was previously believed, the condensation model may require that conditions and timing be "just so" to allow for the rings to survive the processes of planet formation and gas disk dispersal.

## 4.2 Debris from a destroyed satellite?

It was proposed some time ago (Pollack et al., 1973; Pollack 1975; Harris 1984) that Saturn's rings could be the remnant of the catastrophic disruption of one, or several, of Saturn's satellites. Similarly, Uranus's and Neptune's rings are now thought to be the result of the periodic destruction and re-accretion of small moons (see, e.g., Colwell 1994). The "destroyed satellite scenario" requires that a large satellite, already present in Saturn's Roche zone, is destroyed by some mechanism and its fragments scattered away, thus forming a disk like that of Saturn's rings we see today. Such a scenario has two critical points on which some recent results cast new light: (1) how to bring a massive satellite inside Saturn's Roche Zone; and (2) how to destroy it? We now discuss these points.

### 4.2.1 Bringing and keeping a satellite in the Roche Zone

Recent advances in the formation of satellites around gas giant planets suggest possible ways in which satellites may migrate to the planet's Roche zone (Mosqueira and Estrada 2003a,b; Canup and Ward 2006; Estrada et al., 2009). As satellites grow, either by sweep up of dust and rubble followed by gas-drag drift-augmented accretion of smaller satellitesimals and embryos (Cuzzi et al., 1993; Mosqueira and Estrada, 2003a,b), or via binary accretion in a fashion similar to planetary formation (see, e.g., Wetherill 1989; Spaute et al., 1991) occurring under low gas density conditions (Estrada and Mosqueira 2006), several mechanisms can trigger their migration inwards towards the planet. The dominant mechanism for their migration depends on their size, with the transition from gas-drag dominated migration to type-I migration induced by the satellite's tidal interaction with the disk (e.g., Ward 1997) occurring for significantly large bodies (~ 500 km). For example, the inner icy Saturnian satellites remain within the gas-drag dominated to transitional regime (gas drag and torque are similar in importance) for the wide range of subnebula gas surface densities that bracket the models of workers mentioned above (e.g., see Fig. 6 of Mosqueira and Estrada 2003a).

For larger satellites, tidal interactions that trigger type-I migration can lead to their rapid infall towards the planet with some even being lost. In the case of a non-turbulent subnebula, a satellite can migrate into the planet if the satellite's perturbation of the disk is insufficient to stall its migration and open a gap in the gas (Mosqueira and Estrada 2003b). On the other hand, if the subnebula is persistently turbulent, stalling and gap opening are prevented and satellites continue to migrate via type-I. In either case, as the subnebula dissipates or is cleared, eventually both the gas drag and type-I migration timescales become longer than the lifetime of the disk, migration stops, and the surviving satellites remain frozen in their final orbital positions. In this context, then, it is not unlikely that a smaller moon that migrated via gas drag or a larger satellite that migrated via type-I is eventually found near or even inside the planet's Roche Zone (see e.g., Fig. 1 in the Supplementary Online Material of Canup and Ward 2006).

If a satellite is able to migrate and be left stranded in the vicinity of Saturn's Roche zone, a key difficulty then becomes one of keeping the satellite close to Saturn: because of Saturn's tides, a satellite above (resp. below) the synchronous orbit would see its semi-major axis increase (resp. decrease), at a rate given by Murray and Dermott (1999):

$$\frac{da}{dt} = sign(a - a_s) \frac{3 k_{2p} m_s G^{1/2} R_p^5}{Q_p m_p^{1/2} a^{11/2}} \qquad \text{Eq. 21}$$

where a, $m_s$, $m_p$ $R_p$, and G, stand for the satellite's semi-major axis, mass, the planet's mass, the planet's radius, and the gravitational constant, respectively, and $Q_p$ and $k_{2p}$ (~0.3 for Saturn; Dermott et al. 1988) stand for the dissipation factor and the Love number of the planet. However, the value of $Q_p$ that describes all dissipative processes in the planet's interior is very uncertain. Dermott et al. (1988) suggest $Q_p > 1.6 \times 10^4$. In addition it is now thought that $Q_p$ should depend on the excitation frequency; however, this dependence is still not understood or constrained for giant planets. Integrating Eq. 21 yields an ejection timescale of about 1 Gy for a satellite about 3 Mimas mass with $Q_p \sim 10^5$ and ~600 My for satellite with 5 times Mimas' mass. So if the "destroyed satellite" scenario is correct, the destruction must have happened less than 1 Gy after the satellite was brought into Saturn's Roche Zone.

### 4.2.2 Destruction of the satellite

Despite some common ideas, Saturn's tides alone may not be able to destroy and grind a big satellite penetrating its Roche Zone to dust unless the satellite comes close to the planet's surface. The Davidsson (1999) model of tidal fractures shows that a Mimas-sized satellite is disrupted only below 76,000 km from Saturn's center (Saturn's equatorial radius is about 60,300 km for comparison), well inside the C Ring. On the basis of similar arguments, Goldreich and Tremaine (1982) suggest that a 100 km radius satellite can survive undisrupted at 100,000 km from Saturn's center, well inside the B Ring. For these reasons it has been suggested for some time now that only a catastrophic impact with an impactor coming from outside of Saturn's system would be able to destroy such a big satellite and reduce it to a swarm of meter-sized particles (Pollack et al., 1973, Pollack 1975). Because of Saturn's tides, the swarm would not re-accumulate into a single body. Instead, it would then evolve under the effect of dissipative collisions between fragments that lead inexorably to flattening and radial spreading, as for any astrophysical disk (see, e.g., Brahic 1976).

Extrapolating results from hydrocode simulations of catastrophic impacts, we can get an idea of the impactor size necessary to break a parent body (see, e.g., the work of Benz and Asphaug 1999). Assuming an impact velocity ~ 34 km/s (~√3 times the circular orbital velocity at $10^5$ km; Lissauer et al., 1988, Zahnle et al., 2003), breaking an icy body with radius r = 200 km would require an icy impactor between 10 and 20 km radius (Charnoz et al., 2009). The flux of large bodies around Saturn is not well constrained, and extrapolating from the actual visible flux and size distribution of comets in the Solar System (see, e.g., Zahnle et al., 2003) it seems that such an event would be very unlikely in the last $10^8$ years of Solar System history (Harris 1984). Thus both the "destroyed satellite" scenario and the "tidally split comet" scenario (discussed below) face the same problem: the current cometary flux is not sufficient, by orders of magnitude, to provide enough large bodies passing close enough to Saturn (see section 2.2.4).

One possibility is that the breakup of a satellite occurred at the tail end of formation as the subnebula dissipated, or shortly thereafter by a circumsolar interloper. Models of satellite formation suggest that deep in Saturn's potential well, the mid-sized icy moons may have undergone substantial collisional evolution as a result of impacts with incoming planetesimals during their formation process due to a number of factors (Mosqueira and Estrada 2003a; Estrada et al., 2009). However, a problem with this scenario in which a moon is broken up so close to the completion of the satellite system is that it is unclear how much solid material may have been around at the tail end of planet and satellite formation. Presumably, most of the solid mass in Saturn's feeding zone would have been scattered away, but this is not quantified yet.

Another possibility recently proposed by Charnoz et al. (2009) is that a satellite trapped in Saturn's Roche Zone was destroyed much later, during a period called the "Late Heavy Bombardment" (LHB): a short (a few tens of My) global phase of intense bombardment that may have happened throughout the Solar System about 700-800 My after its origin. Whereas the reality of a Solar System-wide bombardment at this time is still a matter of debate, the so-called "Nice Model" (Tsiganis et al., 2005) suggests that the primordial Kuiper Belt, originally 100 to 1000 times more massive than today, was destabilized by Saturn's crossing of the 2:1 mean-motion resonance with Jupiter, triggering a global bombardment in the Solar System. Charnoz et al. (2009) find that a 200 km radius satellite could expect about two destructive impacts during the LHB; i.e., the probability that the satellite would be destroyed is $1 - e^{-2} \sim 90\%$.

The impact scenario during the LHB has some interesting consequences for the uniqueness of massive rings around giant planets: it requires that the ring progenitor remain inside the planet's Roche Zone for 700-800 My in order to be destroyed during the LHB, which is a constraining requirement because of the rapid radial migration induced by the planet's tides. In particular, a satellite below the planet's synchronous orbit would fall rapidly onto the planet, because the migration rate scales as $a^{-11/2}$ (Eq. 21). Thus any planet with its Roche Limit below its synchronous orbit may not be able to keep a large satellite for 800 My inside the Roche Limit. This is precisely the case of Uranus and Neptune, which do not have massive rings like Saturn. Conversely, Saturn's and Jupiter's synchronous orbits are well below their Roche Limits, so that each could maintain a ring progenitor within its Roche Limit at the time of LHB, if it was originally there at the time of the dissipation of their subnebula. Note, however, that the way the quality factor Q depends on the excitation frequency is not well understood for the moment.

Some other major problems remain with the destroyed satellite scenario. In this model, Saturn's rings would be about 3.8-3.9 to 4.5-4.6 Gy old (if formed either during the LHB or during the satellite accretion phase), apparently conflicting with the putative youth of Saturn's rings, unless some mechanism for renewal exists (see sections 2 and 3). In addition, a 200 km moon may contain a substantial fraction of silicates, which is not visible in Saturn's rings (see, e.g., Nicholson et al., 2008). A mechanism to eliminate, or efficiently hide, the silicates still remains to be identified. Alternatively, it may be the case that a broken-up (~ 200 km) moon may have been predominantly icy. Such a notion is not unfounded, given the wide range of densities observed in the Saturnian mid-sized moons. Furthermore, recent models identify and allow for the possibility that there may have been mechanisms at work in the Saturn subnebula that lead to the enhancement of water ice (see section 4.1), especially in the mid-sized satellite region. This is still an open question.

## 4.3 Debris from tidally disrupted comets

Dones (1991) proposed that Saturn's rings could have arisen through the tidal disruption of a large "comet," or outer Solar System planetesimal. A similar "disintegrative capture" model for the origin of the Moon had been proposed by Öpik (1972) and Mitler (1975) prior to the rise of the giant impact theory for lunar origin. The disintegrative capture scenario relies on the difference in gravitational potential across an interloper, i.e., a small body that undergoes a very close encounter (well within the classical Roche limit) with a planet. Consider a small body of radius r and mass $m_p$, that undergoes a parabolic encounter with a planet (i.e., its velocity with respect to the planet at "infinity," $v_\infty$, is equal to zero). Ignore energy dissipation within the body during the encounter. If the body instantaneously breaks into a large number of fragments at its closest approach distance to the planet's center, q, half the fragments (those on the hemisphere facing the planet) will have speeds less than the local escape velocity, $(2GM_{sat}/q)^{1/2}$ for Saturn, and hence will become weakly bound to the planet. For the more realistic case of $v_\infty > 0$, smoothed-particle hydrodynamics simulations, originally carried out for stars tidally disrupted by black holes, indicate that a fraction

$$f = Max\left[\frac{1}{2} - \frac{\left(\frac{1}{2}V_\infty^2 + \frac{Gm_p}{Q}\right)}{1.8\Delta E}, 0\right] \qquad \text{Eq. 22}$$

of the mass of the interloper is captured into orbits around the planet with apocenter distances less than Q. In this expression, $\Delta E = Gm_p r/q^2$ represents the difference in gravitational potential per unit mass across the interloper. The larger $\Delta E$ is, the larger the fraction of material that can be captured. Thus, encounters by large interlopers with small pericenter distances (*q* only slightly greater than the planet's radius) are most effective in this scenario.

Dones (1991) assumed that Chiron, which has $v_\infty = 3$ km/s with respect to Saturn, was a typical Saturn-crosser. At that time, Chiron was the only large Saturn-crosser known. Small bodies with orbits in the Saturn-Neptune region are now known as Centaurs. Even now, only three large Saturn-crossing Centaurs are known (Zahnle et al., 2003; Horner et al., 2004): Chiron, Pholus, and 1995 SN55. Specifically, Dones (1991) carried out Monte Carlo simulations of Centaurs encountering Saturn, using Eq. 22, along with the assumptions that (1) Chiron has a mean radius of 90 km; (2) Centaurs have internal densities of 1 g/cm$^3$; (3) the distribution of $v_\infty$ for Centaurs encountering Saturn is uniform between 0 and 6 km/s; (4) the cumulative size distribution N(r) for Centaurs is a power law with an index of 2, 2.5, or 3 (i.e., N(r) ~ $r^{-2}$, $r^{-2.5}$, or $r^{-3}$); and (5) these size distributions extend up to 250, 500, or 1000 km. Dones estimated that, on average, Centaurs with r > 90 km pass through the "Roche zone" between Saturn's surface and the classical Roche limit once every 28 My. Since the mass of the rings corresponds to (at least) a Mimas-sized body (r = 200 km) with unit density, and at most half of the interloper can be captured (Eq. 22), possible ring progenitors (those capable of depositing the mass of Mimas in a single encounter) pass through the Roche zone once every 200-600 My. However, most of those bodies are moving too fast or do not come close enough to Saturn to leave mass in orbit around the planet.

Dones found that 0.1-1 Mimas masses were captured in 4 Gy at current encounter rates, with most of the capture taking place in rare events by bodies with large r, small q, and small $v_\infty$. He argued that the captured fragments would collide and, by conservation of angular momentum, form a ring near 2 Saturn radii. If the ring progenitor was differentiated, disintegrative capture predicts that material from its icy shell is most easily captured, possibly explaining the icy composition of the rings. However, the rate of ring formation by disintegrative capture is too low in the last billion years to be a likely way to form "young" rings, and there is no obvious reason why Saturn, rather than another giant planet, should have massive rings.

The realization that comet Shoemaker-Levy 9 had been tidally disrupted by Jupiter prior to its fatal descent into the planet in 1994 stimulated efforts to reproduce the "string of pearls" morphology of the comet's fragments. Asphaug and Benz (1996) found that only a strengthless model for the comet, with a pre-disruption diameter of ~1.5 km and a material density of ~ 0.6 g/cm$^3$, could match the observations. Asphaug and Benz also pointed out that because of its low density, Saturn was the giant planet *least* effective in tidally disrupting small bodies.

Dones et al. (2007) revisited the disintegrative capture model, using a modified version of the code that Asphaug and Benz had used to model the disruption of Shoemaker-Levy 9. The main improvement over Dones (1991) was that the disruption and mass capture were modeled explicitly, rather than relying on results in the literature (Fig. 11). Overall, the results of Dones et al. (2007) confirm those of Dones (1991), though it is not known whether Centaurs hundreds of km in size are rubble piles as they assume.

Nonetheless, disintegrative capture remains unlikely to produce a ring system within the last billion years. Dones et al. (2007) used the latest estimates for the population and size distribution of Centaurs, which are generally thought to arise in the Scattered Disk component of the Kuiper Belt. As with large Kuiper Belt Objects, the size distribution of large Centaurs appears to be steep ($N(r) \sim r^{-q}$, with $q \geq 3$), implying that large Centaurs are rare. Since q = 3 was the largest value considered by Dones (1991), and there appear to be only a few Chiron-sized bodies on Saturn-crossing orbits, the rates calculated by Dones (1991) may have been too high. Thus making Saturn's rings recently is difficult, assuming that the current population of Centaurs is representative of the long-term average.

Charnoz et al. (2009) model the influx of planetesimals through the Hill spheres of the giant planets during the Late Heavy Bombardment in the context of the Nice model (see, e.g., Tsiganis et al. 2005). They find that are more than enough interlopers for disintegrative capture to have occurred at each of the giant planets, but find, as in Asphaug and Benz (1996), that Saturn is the planet least likely to tidally disrupt Centaurs. They argue that collisional disruption of a satellite within Saturn's Roche Zone is a better way to explain the origin of the rings, though the rings' pure ice composition is hard to understand in this scenario. This topic is discussed further in Section 4.2.

## 4.4 A conclusion?
After having presented the pros and cons of the different options in detail, is it possible to draw a conclusion about the origin of Saturn's main rings? As shown in sections 2 and 3, for the moment we are still stuck in the longstanding paradox of the apparent youth of the rings, but with no means to form them recently.

Still, the situation is not desperate: our understanding of the origin of other ring systems (those of Uranus, Neptune and Jupiter) has improved, and each time, the collisional origin was the best working scenario, and perhaps it is again the case for Saturn, whereas we still are missing some important observations. Because rings are bombarded with material from all over the Solar System, it seems that, unexpectedly, better understanding the rings' origin requires a better understanding of the full evolution of the Solar System. For example, the recent findings about the Late Heavy Bombardment (Tsiganis et al., 2005), while still controversial, open new possibilities for understanding the rings' origins. Another possibility is that the rings formed just after the accretion of planets (~10-100 My after the origin of the Solar System), as we know the Solar System was dominated by giant collisions between planetary embryos during this epoch. Unfortunately we do not have quantitative measurements about this period.

20 years after the Voyager era, the conclusion that rings are a rapidly evolving system still holds. Perhaps a solution lies in the possibility of recycling material, or that the meteoroid flux is much lower than we have assumed by orders of magnitude, or, finally, that some process somehow slowed down the evolution of the ring system. Since Saturn's main rings are more like a granular medium, whose physics are still poorly understood, we cannot dismiss the fact that we have still a lot to learn about the physics of particulate systems.

So in conclusion, the question of the rings' origin is still open, although formation mechanisms are now better understood.

After having extensively discussed the origin of Saturn's main rings, we now turn to diffuse rings: the F Ring and the E and G Rings. These rings are subject to different processes and their origin and evolution may be somewhat better constrained than those of the main rings.

# 5. Saturn's F Ring: processes and origin

Saturn's F Ring is one of the most dynamic objects in the Solar System. It was first detected by the Pioneer 11 imaging experiment in 1979 (Gehrels et al., 1980). This narrow ring lies 3,400 km beyond the A Ring's outer edge, precisely at the classical Roche limit of Saturn for ice (see section 3). A year later the ring appeared with much greater clarity under the scrutiny of the Voyager 1 cameras, which revealed a remarkable wealth of longitudinal structures, including clumps, kinks, and the so-called "braids" (Smith et al., 1981, 1982). Twenty-five years later, Cassini has provided high resolution images, maps and "movies" of the F Ring, confirming the abundance of delicate dynamical structures. In addition to the visible ring, sharp drops in the flux of magnetospheric electrons detected by Pioneer 11 suggested the presence of a nearby moonlet belt (Cuzzi and Burns 1988). Several moonlets that could be members of this putative belt have been found in Cassini images (Porco et al., 2005; Murray et al., 2005, 2008). The F Ring is also famous for its "shepherding moons" Pandora and Prometheus, which were believed initially to confine the ringlet radially (Goldreich and Tremaine 1982). However, things appear more complex today, and it is not clear at all whether this mechanism is really responsible for the F Ring's narrowness.

All these elements make the F Ring region a complex environment, where the coupling of various physical processes implies a rich evolution.

## 5.1 Characteristics of the F Ring relevant for its origin and evolution

A detailed description of the F Ring is provided in the chapter by Colwell et al., but here we emphasize the aspects relevant for this chapter. The F Ring is an eccentric ringlet, with about 2000 km full radial width but radially sub-divided into a main bright central component (the "core") about 20 km wide in Cassini ISS images, which is surrounded by non-continuous strands of material (Showalter et al., 1992; Murray et al., 1997; Porco et al., 2005). Showalter et al.(1992) found that the ring was 98% fine dust. The outer strands are also composed of dust, and they may form a single spiral structure (Charnoz et al., 2005) due to keplerian shear. These structures seem to be embedded in a continuous sheet of faint material. The core is a moderately high optical depth structure (a few tenths, Showalter et al. 1992), whose orbit seems to be accurately described by the Bosh et al. (2002) model of an eccentric, inclined ringlet precessing under the influence of Saturn's oblateness (Murray et al., 2005). A model of the F Ring occultation data suggests that its opacity is dominated by centimeter-sized particles (Showalter el al. 1992), which may hide within a 1-km wide inner core seen in some recent Cassini images (Murray et al., 2008). Recent occultation data (Esposito et al., 2008a) have also revealed the presence of a population of extended (30 m to 1.5 km) bodies that could be either solid or clumpy aggregates (owing to their translucent appearance). Very high resolution images of the F Ring core (Murray et al., 2008) reveal a wealth of km-scale dynamical structures that still remain to be explained. Conversely, the envelope and strands that surround the core are made of micrometer-sized dust (which is especially bright at high phase angles) with a steep particle size distribution (index of the differential size distribution q~4.6, Showalter 1992). In Showalter (1998, 2004) the largest transient bright features were interpreted as dust clouds generated by meteoroid bombardment, while other authors suggest that local collisional activity involving moonlets (or clumps) could be the cause of these events (Poulet et al., 2000a; Barbara and Esposito 2002; Charnoz 2009). Recent Cassini imaging data seem to support the latter model (Charnoz et al., 2005; Murray et al., 2008; Charnoz 2009). While meteoroid bombardment must also be an active process, it is difficult to quantify for the moment.

The F Ring is much more extended vertically that Saturn's main ring system, and dominates the edge-on brightness during the Sun's crossing of the ring plane ( Nicholson et al., 1996; Poulet et al., 2000a; Charnoz et al., 2001). Photometric models (Poulet et al., 2000b) suggest a vertical extent of about 21 km, thus implying an ongoing dynamical excitation.

The F Ring's mass is unknown. Showalter et al. (1992) obtains a dust mass of $2\text{-}5\times10^9$ kg from the derived size distribution of the envelope, whereas arguments concerning the survival of the core against meteoroid bombardment over the age of the Solar System leads to a core mass estimate equivalent to a 15-70 km radius body, roughly the size of Prometheus or Pandora. The apparent alignment of the F Ring strands with the F Ring core (Murray et al., 1997) suggested that the F Ring core is as massive as Prometheus or Pandora in order to counter-balance the differential precession between the core and the strands. However new models of the strands and jets (Charnoz et al., 2005; Murray et al., 2008) suggest they are transient structures that still require a population of moonlets or clumps to be produced (Murray et al., 2008; Charnoz 2009), thus converging to the same conclusion as Cuzzi and Burns (1988) that suggest that the moonlet population re-accretes the dusty material, thus, extending its lifetime.

In addition to the population of clumps inside the F Ring core found in the UVIS occultations, a population of moonlets, or clumps, exterior to the ring's core has been seen in several ISS images (Porco et al., 2005; Murray et al., 2005, 2008). However, their orbits were difficult to constrain, as the identity of each body is not easily determined due to multiple orbital solutions being possible and the changing appearance of these objects (J. Spitale, N. Cooper, personal communication). The most famous member of this family is S2004/S6 (Porco et al., 2005), whose orbit seems to cross the F Ring core at the precise location where the strands appear to originate (Charnoz et al., 2005).

## 5.2 Processes at work in the F Ring

Such a complex environment should have a rapid dynamical evolution and accordingly, there are strong suspicions that Saturn's F Ring could be young unless a replenishment mechanism exists. Micron-sized particles in the F Ring envelope and strands should leave the ring in far less than $10^6$ years due to Poynting-Robertson drag (Burns et al., 1984). It has been suggested (Showalter 1992) that massive moonlets could maintain the material on horseshoe orbits despite the Poynting-Robertson decay. However this does not seem to be confirmed by numerical simulations (Giuliatti Winter et al., 2004; Giuliatti Winter and Winter 2004). A major factor of dynamical evolution is surely the orbital chaos induced by Prometheus and Pandora, and possibly by the nearby population of moonlets. Since the seminal work of Goldreich and Tremaine (1982), who proposed that the F Ring/Prometheus/Pandora system could be stable due to the so-called "shepherding mechanism" relying on an exchange of angular momentum and collisional dissipation, later analytical studies and numerical simulations (see, e.g., Goldreich and Rappaport 2003a, 2003b; Winter et al., 2007) have shown that the full system could be globally unstable and that orbital chaos is active due to complex interactions between Prometheus with Pandora. Giuliatti Winter et al. (2007) show that moonlets in the F Ring region are rapidly placed on chaotic orbits and scattered over more than 1,000 km in only 160 years, implying that 20 km wide structures like the F Ring core and strands should be young. Charnoz et al. (2005) find that the F Ring strands should disappear in about 1800 orbits, only ~3 years, for the same reasons. In addition, the evolution of Pandora and Prometheus's semi-major axes due to tidal interactions with Saturn's rings should lead to an encounter of Prometheus with the F Ring in less than $10^7$ years (Poulet and Sicardy 2001). Defying all these arguments, the presence of a narrow, uniformly precessing core within such a chaotic region remains unexplained (Bosh et al., 2002; Murray et al., 2008). Due to meteoroid bombardment, big bodies should erode by ~ $3 \times 10^{-5}$ cm/year (Showalter 1992), so that bodies smaller than 100 m should be much younger than the Solar System. All these destruction mechanisms are tempered by the possibility that substantial accretion could be expected in Saturn's F Ring (see section 2.2.4). Barbara and Esposito (2002) show that the competition between accretion and fragmentation should result in a bi-modal size distribution. While such a distribution has not been observed in UVIS data (Esposito et al., 2008a) we note that the average number of bodies larger than 1 km diameter seems to be correctly predicted by this model (Barbara and Esposito 2002).

**5.3 Origin and evolution of Saturn's F Ring**
It is not an easy task to draw a coherent picture of the F Ring's origin with such a diversity of antagonistic mechanisms. However, we note that different models have been proposed with a similar underlying principle: the progressive erosion of a population of big bodies by a diversity of mechanisms such as meteoroid bombardment and collisional evolution.

Cuzzi and Burns (1988) proposed that the Saturn's F ring dusty component is a cloud of particles released by the collisional erosion of an underlying population of 0.1-1 km sized moonlets. They developed a detailed model in which moonlets collide and release surface material that give birth to the F Ring itself. The material is subsequently re-accreted by the same moonlets. They computed that a population of 100 m-sized moonlets must represent a total optical depth about $10^{-4}$ (Cuzzi and Burns 1988). This model is very similar to the modern picture of a circumstellar debris disk like Beta Pictoris (see, e.g., Lagage and Pantin 1994), in which unseen belts of small bodies, stirred by one or several planets, produce a dusty disk that is visible thanks to its infrared excess. Showalter (1992) argued such a large population of moonlets is not compatible with Voyager images and also that, since the F ring core seems to be only ~1 km wide (recently confirmed in Murray et al., 2008), encounter velocities between moonlets would be too low (~0.1 m/s) to release material over hundred of kilometers. In addition, Voyager data revealed the presence of some clumps suddenly appearing and dissipating. Showalter (1992) proposed that the population of clumps and moonlets embedded in the core is ground to dust via meteoritic bombardment impacting the ring at high velocity (~30 km/s), as in other dusty rings of the Solar System. Note that the Showalter (1992) model is not incompatible with the presence of moonlets suggested by Cuzzi and Burns (1988); it merely proposes an alternative mechanism for dust production. After the Cassini high-resolution images of the F Ring and the discovery of the population of nearby moonlets (Porco et al., 2005), it was proposed (Charnoz et al., 2005, Murray et al., 2008, Charnoz 2009) that the population of moonlets exterior to the core (among which S2004/S6 is a member) regularly collides with the population inside the core, releasing material whose orbital motion form structures described as "spirals" (Charnoz et al., 2005) or "jets" when they are still young (Murray et al., 2008; Charnoz 2009). In this view (close to the Cuzzi and Burns model), these two populations interact physically and gravitationally and create the transient and dusty structures surrounding the F Ring.
So it seems that the F Ring's origin is linked to the origin of the population of km-sized moonlets. Cuzzi and Burns (1988) proposed that one or several small moons were destroyed in the past, whose fragments are slowly eroding today. Conversely, Barbara and Esposito (2002) suggest that there is on-going accretion in the F Ring core, regularly producing clumps and moonlets, whose subsequent collisional erosion produces the F Ring. UVIS stellar occultation data (Esposito et al., 2008a) have revealed the presence of a population of clumps in the core, among which some are translucent, like loose gravitational aggregates, suggesting that that accretion is active in the densest regions of the F Ring (see section 2.3 and Fig. 2). However, none of these models at the moment can be reconciled with the "hot" structure of the ring (i.e., the large random velocities) induced by the strong nearby perturbations (see chapter by Colwell et al.) with the confirmed presence of a narrow and thin core that clearly must be a "cold" structure (Showalter 2004; Murray et al. 2008). Is self-gravity the solution? We have no clue for the moment.

# 6. Diffuse rings: processes and origins

After having discussed the origin and evolution of Saturn's F Ring, we now turn to the origin and evolution of the E and G Rings. While they share similarities with the F Ring in terms of their optical properties, the E and G Rings may have a different origin and fate due to their erosional origins and the importance of non-gravitational forces and resonances in their dynamics. The chapter by Horanyi et al. provides a detailed discussion of these rings, while here we briefly review Cassini's contributions to our understanding of how the E and G Rings are sustained.

The E and G Rings, like the D and F Rings and a number of other structures, can be distinguished from Saturn's better known main rings not only by their low optical depths, but also by the fact that they dramatically increase in brightness at high phase angles. All these rings therefore must contain a sizable fraction of particles that are sufficiently small to efficiently scatter light by as much as several tens of degrees via diffraction. Such particles are of order 1-10 microns in size, and are therefore extremely sensitive to non-gravitational processes that can both erode and disperse material on relatively short timescales (Burns et al., 2001). For example, sputtering by energetic particles in Saturn's magnetosphere can destroy particles of this size in a few thousand years, while charge variations can potentially cause orbital evolution in even less time. The visible dust therefore almost certainly needs to be continuously re-supplied to these rings by various sources.

The source of the most extensive of the dusty rings, the E Ring, was suspected to be the satellite Enceladus long before Cassini reached Saturn. After all, Enceladus' orbit is located near the peak in the E Ring's brightness. However, prior to Cassini's arrival it was not clear how Enceladus generated the E Ring. The blue color of this ring in backscattered light observed by Earth-based telescopes, which strongly contrasts with the neutral or red color of many other dusty rings (Nicholson et al., 1996; de Pater et al., 1996; Bauer et al., 1997; de Pater et al., 2004) suggested that this ring had an unusually narrow or steep particle size distribution (Showalter et al., 1991), such that one-micron particles were emphasized. Such an unusual size distribution can be explained if non-gravitational forces are involved in generating and/or dispersing material in this ring. For example, Hamilton and Burns (1994) developed a model in which the E Ring was self-sustained by the impacts of small grains in the ring onto various satellites. In this scenario, solar radiation pressure and electromagnetic interactions between the particles in the ring and the surrounding plasma preferentially gives particles of a certain size large eccentricities, and so only particles of that size would be able to populate an extensive ring or to impact moons with enough speed to yield additional E Ring material. One issue with such models was that Enceladus has a relatively high escape speed, and yet it was not obvious why its particle yield was sufficiently large to support the E Ring.

Cassini images finally showed how Enceladus is able to supply the E Ring: its south pole is geologically active, generating a plume of small particles that extends thousands of kilometers above its surface (Porco et al., 2006). While the physical processes responsible for generating this plume are still being debated (Porco et al., 2006, Kieffer et al., 2006, Kieffer and Jakosky 2008), the connection between the plume and the E Ring is secure. Both contain ice-rich particles a few microns across (Kempf et al., 2008; Postberg et al., 2008; Hedman et al., 2009b; Briallianov et al., 2008). In fact, the dynamics of the plume particles can provide a natural explanation for the E Ring's peculiar size distribution. For the plume particles to populate the E Ring, they must be launched from the surface with sufficient velocity to escape Enceladus' gravity. Physical models of the condensation and acceleration of solid matter within cracks indicate that larger grains reach the surface with lower average velocities (Schmidt et al., 2008), and so the size distribution of particles that make it into the E Ring is expected to be skewed towards smaller particles. Detailed analyses of the plume's spectral and photometric properties have recently confirmed that different sized particles are launched from Enceladus with different velocity distributions (Hedman et al., 2009b), and further studies of remote-sensing and in-situ measurements should provide additional constraints on the dynamics of how the plume supplies the E Ring. For more information about how the particles supplied by Enceladus are dispersed throughout the E Ring, see the chapter by Horanyi et al.

In contrast to the E Ring, the origins of the G Ring were obscure before Cassini. The limited in-situ measurements and the observed spectral and photometric properties of this ring were consistent with models of collisional debris (Showalter 1993; Throop and Esposito 1998; de Pater et al., 2004). The visible dust in this ring therefore appeared to be generated by collisions among and into a suite of larger (>1 m) bodies located in the region of the G Ring (Canup and Esposito 1997). However the G Ring's brightness peaks around 168,000 km from Saturn's center, over 15,000 km from the nearest known satellite. There was no clear explanation why source bodies would be concentrated in this particular location.

A combination of in-situ and remote-sensing data from the Cassini spacecraft has clarified the source of the G Ring (Hedman et al., 2007a). Images taken by the cameras reveal a localized brightness enhancement at 167,500 km from Saturn's center, near the inner edge of the G ring. This arc has a radial full-width at half-maximum of ~250 km and extends over only $60^o$ in longitude. Cassini has imaged this arc multiple times over the course of the nominal mission, and these data show that the arc moves around Saturn at a rate consistent with the 7:6 corotation eccentricity resonance (CER) with Mimas. Numerical simulations confirm that the gravitational perturbations due to this resonance are able to confine material into an arc.

In-situ measurements of the charged particle environment by the MIMI instrument demonstrated that this arc does not just contain the dust visible in most images. MIMI detected a ~50% reduction in the energetic electron flux when it passed through magnetic field lines that thread through the arc. Such deep absorptions were not observed on other occasions when the spacecraft flew over longitudes in the G Ring far from the arc, and the radial width of this absorption was comparable to the width of the arc observed in images, so this absorption can reasonably be attributed to material trapped in the arc. The magnitude of the absorption indicates that this arc contains a total mass between $10^8$ and $10^{10}$ kilograms. This greatly exceeds the observed amount of dust, and is therefore direct evidence that larger bodies exist in the arc. The existence of at least one such object has since been confirmed, thanks to images that appear to show a small moonlet embedded within the arc (Porco et al., 2009).

Based on the above information, Hedman et al. (2007a) have suggested that the arc contains a population of large particles that may be the remnants of a disrupted moonlet. The 7:6 Mimas CER prevents the material from dispersing and therefore helps explain why a relatively dense ring exists at this location. Collisions into and among these larger bodies produce the dust that forms the visible arc. Unlike larger objects, these small dust grains are subject to significant non-gravitational forces. In particular, interactions between the dust grains and the local plasma (which co-rotates with Saturn's magnetosphere) can accelerate these small particles, enabling them to leak out of the arc and drift outward to form the rest of the G Ring. As the dust grains drift away from Saturn, processes like sputtering and micrometeoroid bombardment steadily erode them, causing the density and brightness of the ring to steadily decline with radial distance from the arc. Now that ultra-faint rings and resonantly-confined arcs of debris have been found associated with several small moons of Saturn (Hedman et al. 2009a), comparisons among these different moon-ring systems should allow such models to be more thoroughly tested and therefore yield a better understanding of the origins and evolution of the G Ring.

# 7. Conclusions

Thanks to Cassini findings, we now seem to have quite a clear picture of the origin of the E and G Rings: the E Ring is fed by Enceladus' plumes and the progressive destruction of a moonlet feeds the G Ring. The question of the F Ring's origin is still unsolved, but the discovery of a population of nearby moonlets, as well as high resolution images of the core, seem to qualitatively confirm the model of Cuzzi and Burns (1988), according to which the F Ring is the result of the collisional evolution of a moonlet belt, whose origin is still to be understood.

Conversely, the question of the origin of Saturn's main rings is still wide open more than 25 years after the flybys of Voyager 1 and 2. Since then, new observational data and theoretical results have brought new insights both into ring dynamics and into the history of the Solar System, opening new possibilities for understanding the main rings' origin and long term evolution. These developments include new models of the formation of giant planets and their satellites, local numerical simulations of planetary rings, and the discovery of the "propellers" in the A Ring. The Voyager-era idea of young rings created recently by disruption of a Mimas-size moon (e.g., Esposito 1986) seems untenable, since the likelihood of such an event in the last 100 Myr is tiny. Still, no fully satisfactory answer has been found, and the apparent contradiction between the rings' apparent youth and the difficulty of forming them in the last billion years still holds.

One of the strongest constraints on the rings' age – the darkening by meteoroid bombardment – might be solved if the rings were more massive than previously thought or, alternatively, if the meteoroid bombardment was much smaller than previously estimated. However, as mentioned in section 3.1, the meteoroid flux may also have been much higher in the distant past. The idea of recycling material, embodied in the concept of "Cosmic Recycling" still needs modeling and investigation. The evidence for ancient rings and continued recycling is indirect, but more than just intriguing. It includes the larger optical depths measured by Cassini occultations, the clumpiness seen in the rings that implies that the mass of the rings may have been underestimated, and the variety of structures visible in the outer A Ring and in the F Ring.

Concerning the mechanism responsible for the implantation of the ring system, new theoretical results on the Late Heavy Bombardment open the possibility that either the destruction of a satellite or the tidal splitting of passing comets could have taken place about 700-800 My after the origin of the Solar System. If either of these is the correct explanation for the origin of the rings, the rings would be about 3.8-3.9 billion years old, still too old given the pollution and evolutionary processes at work in the rings, and a recycling mechanism would still be necessary.

Data that would help further progress on the question of rings' origins are:

1- The mass of Saturn's main rings, in particular for the B Ring.
2- The meteoroid flux at Saturn. This would give almost an absolute measurement of the rings' age by at least two independent methods (structural and photometric appearances).
3- Images and spectra of individual particles, their size distribution and how it depends on the distance to Saturn. Such data would help to better constrain evolutionary processes and in turn, the rings' origin. In addition, we still do not know whether the silicates are really absent from Saturn's rings or if they are deeply buried inside the ring particles. An answer to this question would be critical and would provide strong constraints on the scenario of rings formation, since, we do not know of an obvious way to eliminate silicates from Saturn's rings.

Data #1 and #2 could be provided during the Cassini extended-extended mission after 2010, whereas datum #3 could be only obtained with an in situ mission in Saturn's rings (see the chapter by Guillot et al.). A final important source of data that may be available in the moderately near future would be the occurrence of ring systems around extra-solar planets, which would allow us to test whether massive rings are either a natural or an extraordinary outcome of planetary formation. These data would also be invaluable for building a scenario for the origins of Saturn's main rings. The Corot and Kepler missions will hopefully bring new constraints about the presence of rings around extra-solar giant planets, and in turn, give new insight into the origin of Saturn's rings.

**Acknowledgements:** We wish to thank N. Albers, M. Showalter and K. Ohtsuki for valuable comments. We also thank our three referees, whose comments improved the quality of this chapter.

# TABLES

| | Location[1] (Width) | Optical Depth | Dust Fraction (%) | Size power-Law Index[1] | Notes |
|---|---|---|---|---|---|
| D ring | 66,000– 74,000 | $10^{-3}$ | 50-100 | ? | Internal structure |
| C ring | 74,490– 91,983 | ~ 0.1 | <3 | 3.1 | Some isolated ringlets |
| B ring | 91,983–117,516 | 1-5 | <3 | 2.75 | Abundant structure |
| Cassini Division | 117,516– 122,053 | 0.05-0.15 | <3 | | Several plateaus |
| A ring | 122,053– 136,774 | ~ 0.5 | <3 | a<10m [3] 2.75-2.90 a>10m [3] 5-10 | Many density waves Propellers[3] |
| F ring | 140,200 (W≅50 km) | 0.1-0.5 | >98 | 2-3 | Narrow, broad components |
| G ring | 166,000– 173,000 | $10^{-6}$ | >99 | 1.5-3.5 | |
| E ring | 180,000– 450,000 | $10^{-5}$ | 100 | | Peak near Enceladus |

**Table 1:** Saturn's rings. Sources: Burns et al., 2001, Nicholson and Dones, 1991. For more detail, see Chapter 14 by Colwell et al. (1) Distance units in km (2) power law index of the differential size distribution (3) Tiscareno et al.(2006) and Sremčević et al.(2007) show that there is a knee in the size distribution around a=10m. See section 2.1.1 for details.

| Ring Feature | Inferred/ observed age | Notes | OLD | YOUNG | RENEWED |
|---|---|---|---|---|---|
| Narrow ringlets in gaps | months | Variable during Cassini mission | | **OK** | **OK** |
| F ring clumps | months | Sizes not a collisional distribution | | | **OK** |
| F ring moonlets | tens to millions of years | Create fans and jets | | **OK** | **OK** |
| Cassini Division density waves | 100,000 years | Low mass quickly ground to dust | **NO** | **OK** | |
| Embedded moonlets | millions of years | Low bulk density shows accretion | | **OK** | **OK** |
| "Propeller" objects | millions of years | Steep size distribution from recent disruption | **NO** | **OK** | |
| Pollution and Color of A ring | $10^7 - 10^8$ years | Expected more polluted than B | **NO** | **OK** | **NO** |
| Pollution and Color of B ring | $10^8 - 10^9$ years | Meteoroid flux not so high? More massive? | **NO** | **OK** | **OK if massive** |
| Color/spectrum varies in A | $10^6 - 10^7$ years | Ring composition not homogenized | | | **OK** |
| Shepherd moons | Breakup: $10^7$ years | | | **OK** | **OK** |
| Radial spreading | Momentum: $10^7$ years | Breakup/momentum: No contradiction in ages! | **NO** | **OK** | |
| Self-gravity wakes | days | Particles continually collide; self-gravity and adhesion enhance aggregation | **OK** | **OK** | **OK** |

**Table 2.** Inferred ages of various ring features and consistency with 3 models for ring formation. OK: Can be accommodated; NO: Serious contradiction; Blank: Unclear, or deserves more study. Adapted from Esposito 2006, presented at the Montana Rings Workshop.

# FIGURES

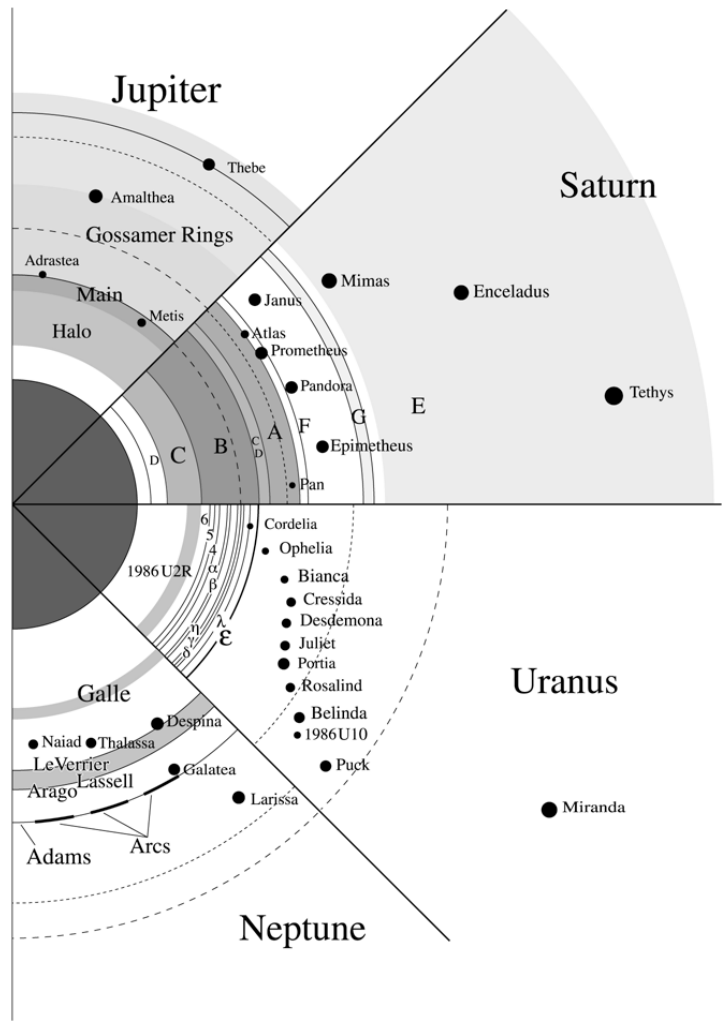

**Figure. 1.** A comparison of the four planetary ring systems, including the nearby satellites, scaled to a common planetary equatorial radius. Density of crosshatching indicates the relative optical depth of the different ring components. Synchronous orbit is indicated by a dashed line, the Roche limit for a density of 1 gm cm$^{-3}$ by a dot-dash line. Figure courtesy of Judith K. Burns, from Burns, et al. (2001).

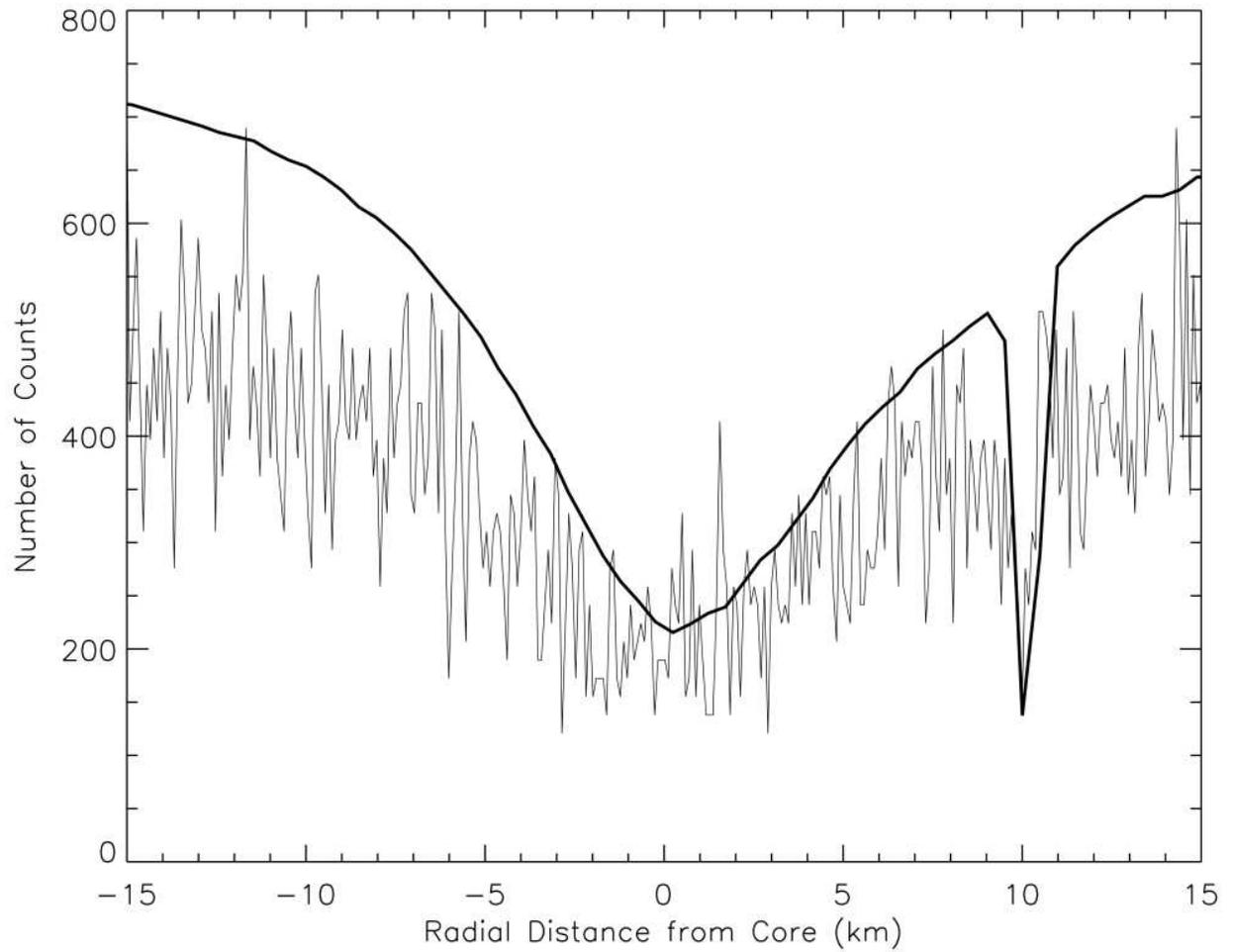

**Figure 2**. VIMS (solid, smooth black curve) and UVIS (thin, gray curve) alpha Sco Egress occultation data overplotted. The UVIS data curve (scaled to match VIMS unocculted flux far from the event, off the figure) appears noisier, but at the center of the event has higher spatial resolution. Pywacket, the event 10 km outside the F ring core, is detected by both instruments.

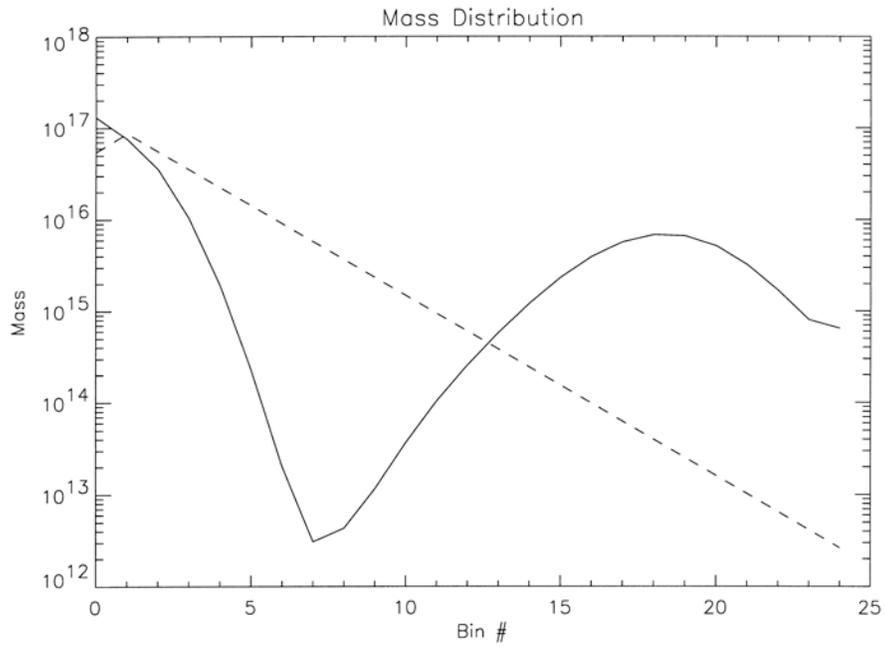

**Figure 3**. Initial and evolved mass distribution for Saturn's F ring. Initially a shattered moon gives a power-law size distribution (dashed line). The system evolves to a bi-modal distribution (solid line) due to particle interactions and accretion. The size bins run from dust to small moons. Barbara and Esposito (2002).

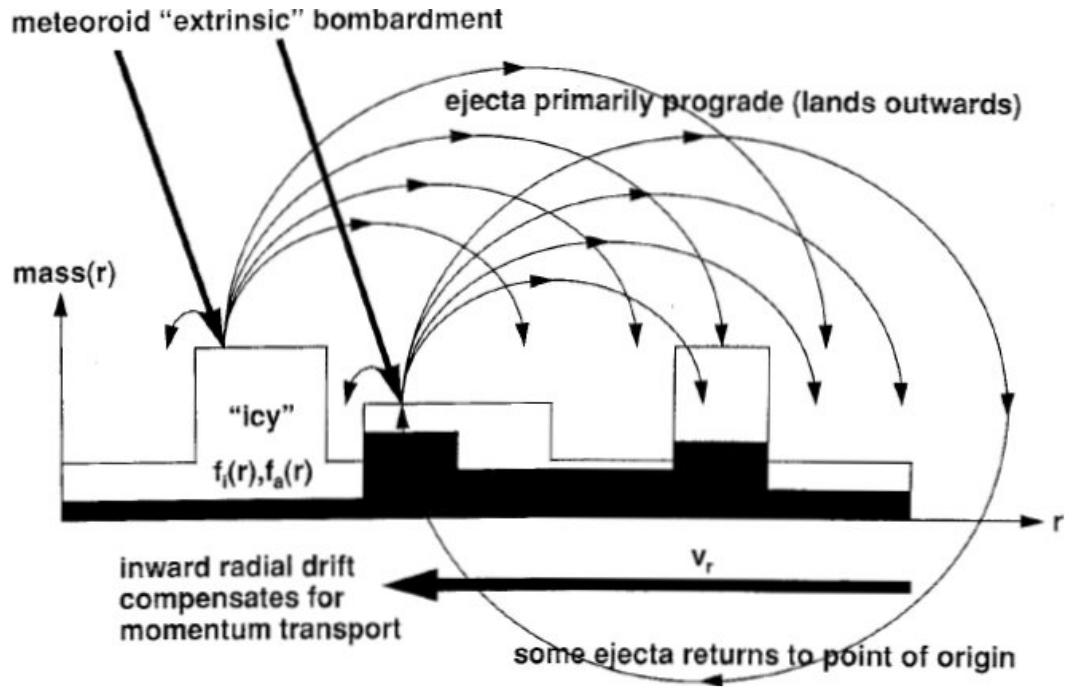

**Figure 4:** Schematic of the ballistic and subsequent pollution transport process. Impact ejecta (which are primarily prograde) carry both mass and angular momentum to outer regions of the rings where the specific angular momentum is typically larger. This leads to a net inward radial drift of material where ejecta preferentially land which compensates for the direct momentum transport with respect to drift due to the rings' inherent viscosity. The various plateaus represent radial distribution of mass density, while shaded regions represent the variable fraction of material that is non-icy (From Cuzzi and Estrada, 1998).

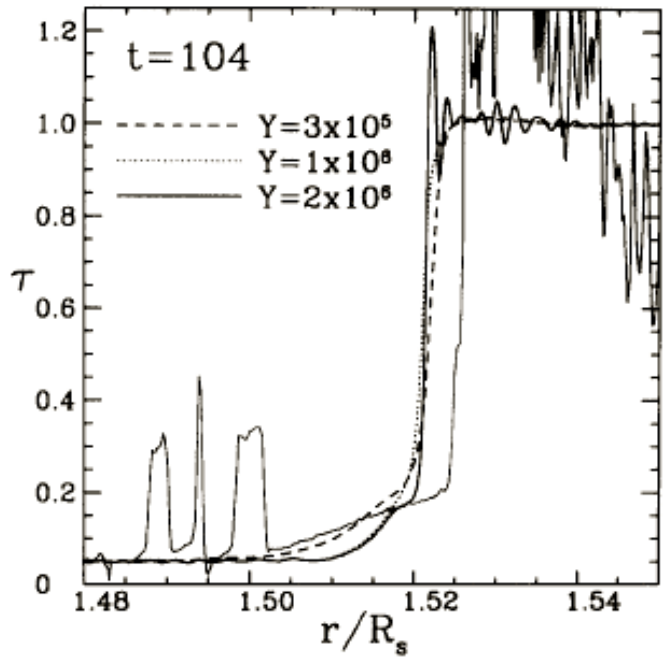

**Figure 5**: A comparison of the Voyager ISS optical depth (light solid curve) with ballistic transport simulation results using a power-law distribution of ejecta velocities and different values for the total yield Y. The simulations were run for 104 "gross erosion times," which refers to the time a reference ring annulus would disappear due to ejected material if nothing returned. The lower bound for the ejecta distribution is varied, but is typically ~ 1 m s$^{-1}$. The undulatory structure discussed in the text in the inner B ring is clearly seen. For details, see Durisen et al. (1992).

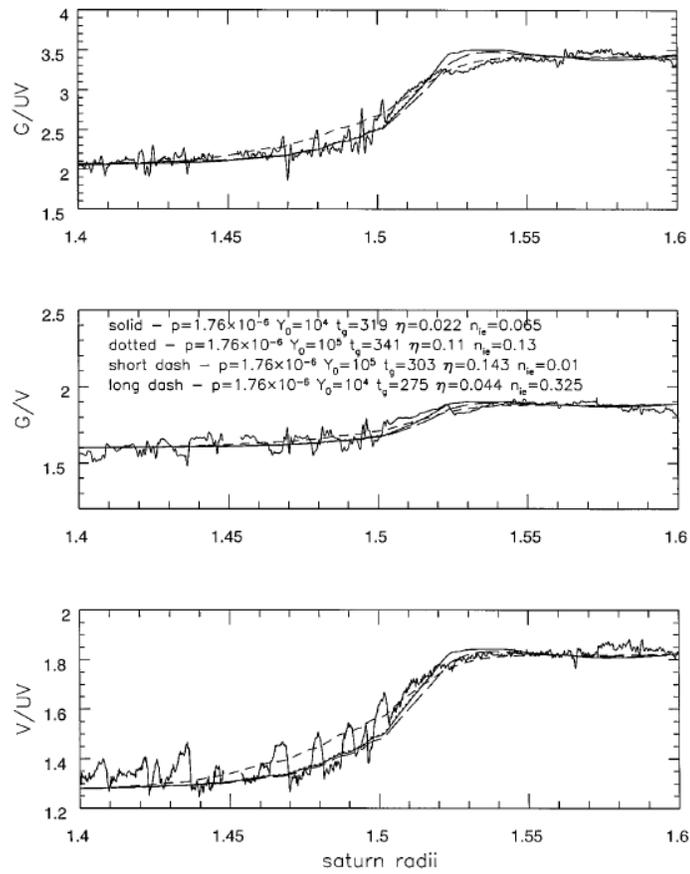

**Figure 6**: Radial profiles of Voyager G/UV, V/UV, and G/V color ratios, compared with the results of combined ballistic transport and radiative transfer models. Although the absolute age of the rings depend on the micrometeoroid flux, the retention efficiency η, and the impact yield, the shape of the B Ring/C Ting transition depends on a combination of these parameters embodied by the shape parameter p. The figure illustrates how all three full radial profiles of color ratio can be simultaneously matched. The best case fit (solid curve) corresponds to $t_G \sim 300$ "gross erosion times" or a timescale of $\sim 3\text{-}4 \times 10^8$ years. In the figure, $Y_0$ is the ejecta yield, and $n_{ie}$ is the imaginary component of the refractive index of extrinsic material.

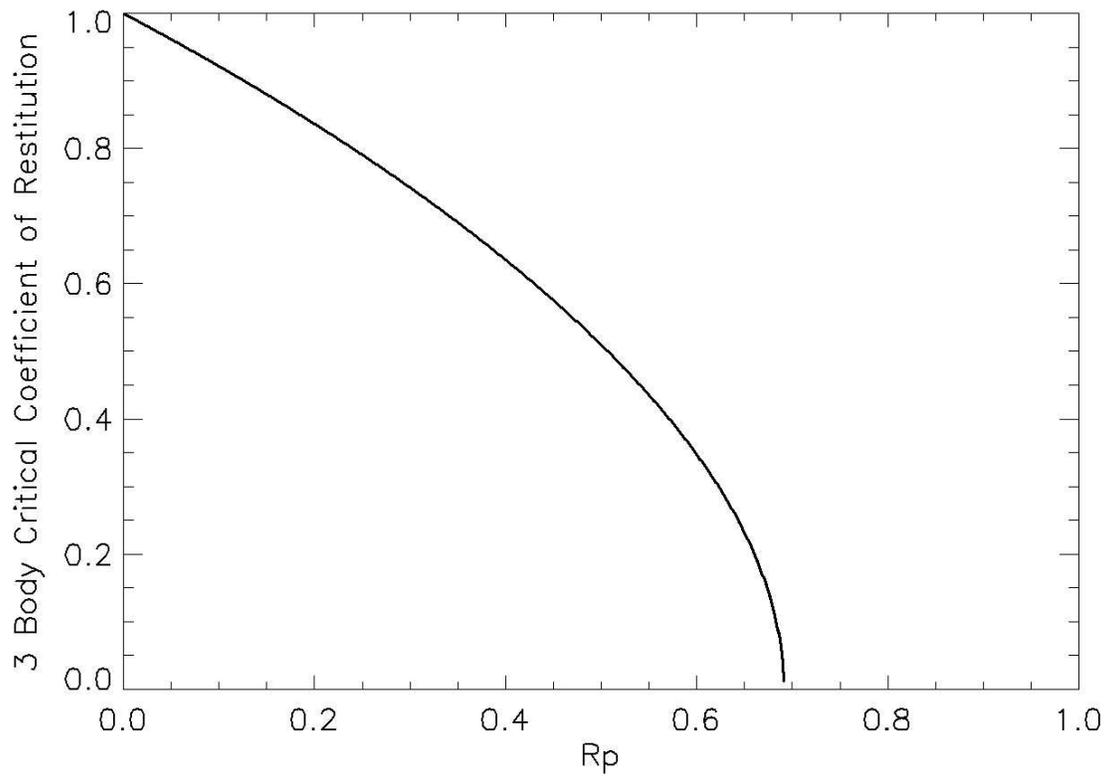

 **Figures 7**: Critical coefficient of restitution for the classical three body gravitational encounter (from Canup and Esposito 1995).

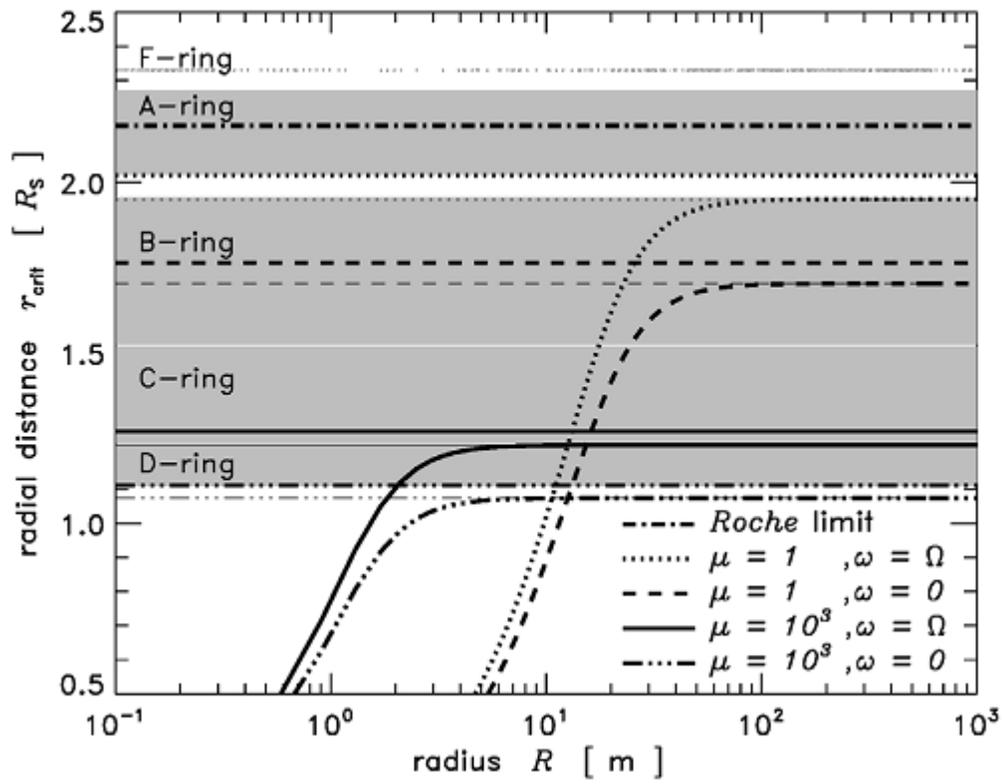

**Figure 8**: Critical distance $R_{crit}$ as a function of particle radius R and particle size ratio $\mu$. An agglomerate combination is stable in the regime above/left of the corresponding line. Any combination below/right of is unstable (From Albers & Spahn 2006).

**Figure 9.** In the collisional cascade, moons are shattered and their fragments further broken to make rings and dust. Eventually, the last moon is destroyed and the original material is completely ground to dust (from Esposito and Colwell, 2003).

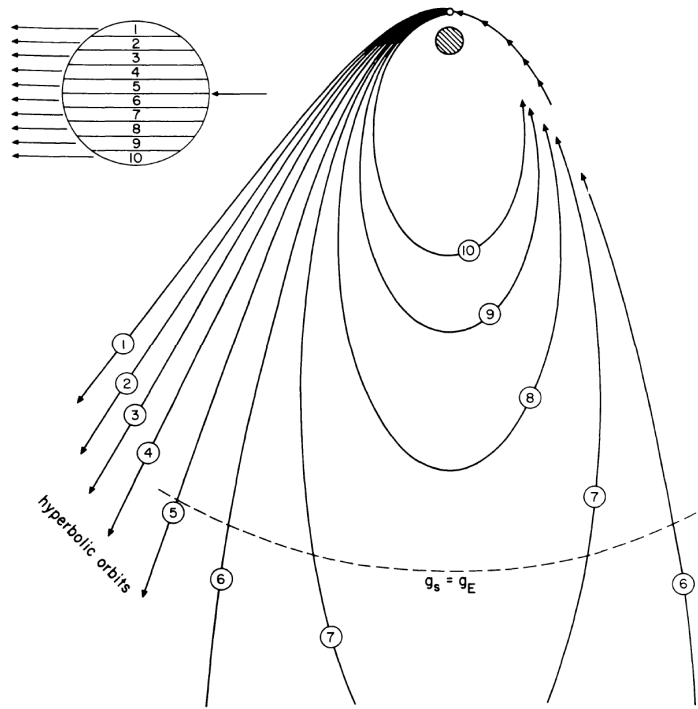

**Figure 10.** Diagram illustrating the process of disintegrative capture, from Wood and Mitler (1974) and Wood (1986).

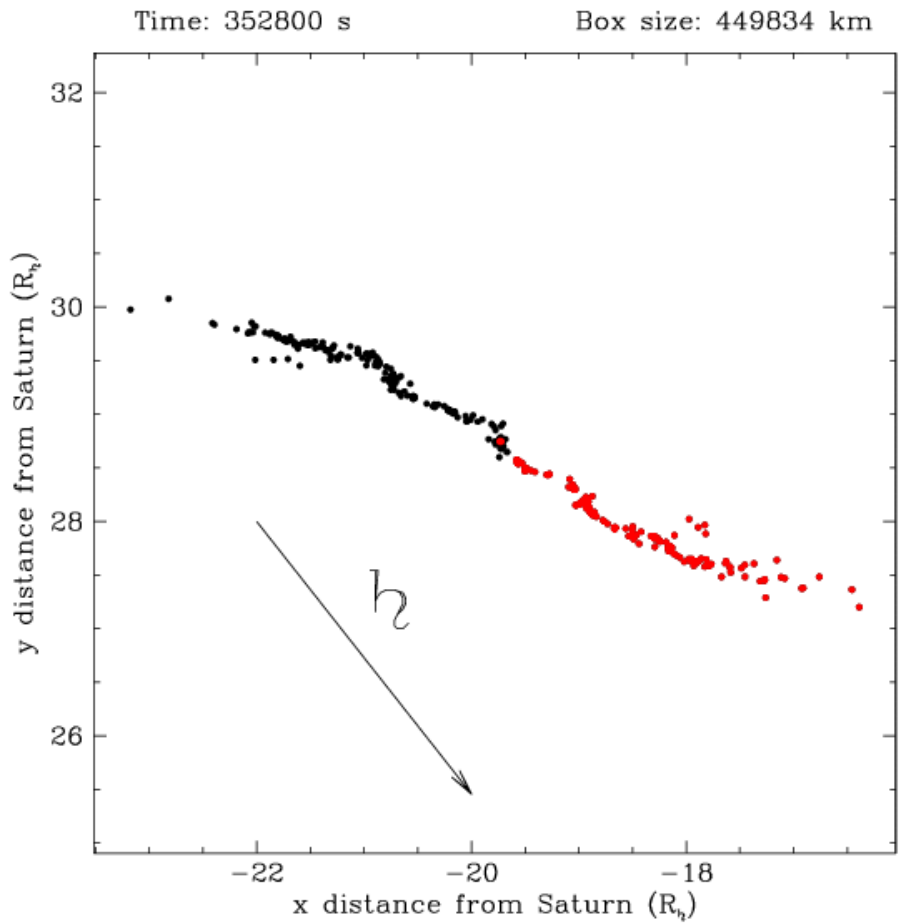

**Figure 11**. Panel from simulations by Dones et al. (2007) of a large Centaur tidally disrupted by Saturn after it passed within 1.1 radii of the planet's center. Several days after the encounter, the fragments are beyond Titan's orbit. The black dots represent fragments that will escape the Saturn system. The red dots represent fragments bound to Saturn; these fragments were on the hemisphere of the Centaur facing Saturn at the time of disruption. About 40% of the Centaur's mass was captured. The objects on bound orbits have semi-major axes around Saturn of hundreds of planetary radii, in the region of Saturn's irregular satellites. Plausibly, collisions between the fragments will ultimately result in a ring within Saturn's Roche Zone, but this process has not yet been modeled.